\documentclass[a4paper,onecolumn,11pt,unpublished]{quantumarticle}
\pdfoutput=1
\usepackage[utf8]{inputenc}
\usepackage[english]{babel}
\usepackage[T1]{fontenc}
\usepackage{amsmath}
\usepackage{amsthm}
\usepackage{amssymb}
\usepackage{hyperref}
\usepackage{braket}
\usepackage{tikz}
\usepackage{quantikz}
\usepackage{float}
\usepackage{pgfplots}
\usepackage{graphicx} 
\usepackage{caption}
\usepackage{subfig}
\usepackage[sorting=none,url=false,minnames=1,maxnames=1,date=year,style=numeric-comp]{biblatex}
\DeclareFieldFormat*{title}{\mkbibquote{#1\isdot}}
\usepackage{svg}
\usepackage{xcolor}
\usepackage{enumitem}
\usepackage{makecell}
\hypersetup{
    citecolor=quantumviolet,
}

\newtheorem{theorem}{Theorem}
\newtheorem{lemma}{Lemma}

\newtheorem{remark}{Remark}

\newcommand{\Rop}{R}
\newcommand{\bit}[1]{\texttt{#1}}
\newcommand{\scalingFactorBE}{\lambda}
\newcommand{\opSelect}{\ifmmode\text{\textsc{Select}}\else\textsc{Select}\fi}
\newcommand{\opPrepare}{\ifmmode\text{\textsc{Prepare}}\else\textsc{Prepare}\fi}
\newcommand{\opTag}{\ifmmode\text{\textsc{Tag}}\else\textsc{Tag}\fi}
\newcommand{\opRestore}{\ifmmode\text{\textsc{Restore}}\else\textsc{Restore}\fi}

\begin{document}

\title{TARE: Block Encoding Linear Combinations of Pauli Strings Without Ancilla State Preparation}
\author{Niclas Schillo}
\email{niclas.schillo@iao.fraunhofer.de}
\affiliation{%
    Fraunhofer-Institut für Arbeitswirtschaft und Organisation IAO,
	Nobelstra{\ss}e 12,
	70569 Stuttgart,
	Germany}
\affiliation{%
    Universit\"at Stuttgart, Institut für Arbeitswissenschaft und Technologiemanagement IAT,
    Nobelstra{\ss}e 12,
    70569 Stuttgart,
    Germany}
\author{Andreas Sturm}
\affiliation{%
    Fraunhofer-Institut für Arbeitswirtschaft und Organisation IAO,
	Nobelstra{\ss}e 12,
	70569 Stuttgart,
	Germany}
\email{andreas.sturm@iao.fraunhofer.de}
\author{Rüdiger Quay}
\affiliation{%
    Fraunhofer-Institut für Angewandte Festkörperphysik IAF,
	Tullastra{\ss}e 72,
	79108 Freiburg, Germany}
\affiliation{Albert-Ludwigs-Universität Freiburg, Fritz-Hüttinger Chair for Energy-Efficient High-Frequency Electronics EEH,
    Emmy-Noether-Straße 2,
    79110 Freiburg, Germany}

\begin{abstract}
Quantum algorithms based on Quantum Signal Processing (QSP) offer the potential for speedups across a broad range of applications, with block encodings serving as the central input model. In this framework, non-unitary matrices are embedded into larger unitary operators, and the cost of constructing these encodings often dominates the overall gate complexity. In this work, we introduce Tag-and-Restore Encoding (TARE), a block-encoding method for linear combinations of Pauli strings. In this method coefficient magnitudes are absorbed into a unitary built from a set of mutually anti-commuting Pauli strings acting on the system register. These Pauli strings are then mapped to the target Pauli strings through appropriate transformations, yielding a block encoding of the target operator. The ancilla register size scales logarithmically with the number of Pauli strings and can be extended to larger registers providing a width/depth tradeoff. We evaluate TARE through numerical simulations of the transverse-field Ising model, the Jordan-Wigner image of a fermionic star Hamiltonian, and random Pauli-string operators. Compared with standard Linear Combination of Unitaries (LCU), TARE substantially reduces the T-gate count while improving circuit depth in several cases. These results suggest that TARE can provide resource-efficient block encodings for a wide range of relevant systems.
\end{abstract}
\section{Introduction} 
Block encoding is an essential subroutine in quantum algorithms based on quantum signal processing \cite{Low_2017, Low_2019}. These include the Quantum Eigenvalue Transformation (QET), the Quantum Singular Value Transformation (QSVT) \cite{Gily_n_2019, Martyn_2021}, and their generalizations \cite{PRXQuantum.5.020368, sünderhauf2023generalizedquantumsingularvalue}, which solve a range of numerical linear algebra problems on quantum computers. At its core, block encoding embeds a non-unitary matrix as a block of a larger unitary matrix using ancilla qubits and post-selection. Since quantum circuits can implement only unitary operations, block encoding is the standard way to apply non-unitary operators on a quantum computer.\\
The most prominent block-encoding scheme is the Linear Combination of Unitaries (LCU) method \cite{Childs_2012}, which implements non-unitary operators as weighted sums of unitary matrices. A central component of LCU is the \opPrepare{} oracle that loads the operator coefficients onto an ancilla register, and reducing its cost has motivated a range of constructions, including QROM- and QROAM-based coefficient loading \cite{Babbush_2018_qrom, Berry_2019_qroam} and Dicke-state-based preparations \cite{dellachiara2025efficientlcublockencodings}. Further LCU variants reduce the overall cost through alternative implementation architectures \cite{Chakraborty_2024}, parallelisation of the \opSelect{} subroutine via commuting operators \cite{boyd2024lowoverheadparallelisationlcu}, and unifying decomposition frameworks for fermionic Hamiltonians \cite{Loaiza_2025_Majorana}. Beyond LCU, several other block-encoding schemes have been developed to address specific matrix structures and resource constraints. These include explicit block encodings of structured matrices \cite{doi:10.1137/22M1484298, S_nderhauf_2024, sturm2025efficientexplicitblockencoding, liu2024efficientquantumcircuitblock, Camps_2022_FABLE}, hardware-efficient techniques such as Hamiltonian embedding for noisy intermediate-scale quantum (NISQ) devices \cite{Leng_2025}, and variational block-encoding methods \cite{Kikuchi_2023, rullkötter2025resourceefficientvariationalblockencoding} that optimize encoding parameters through classical simulation.\\
Despite the potential quantum speedups of QSVT-type algorithms, their practical performance is often limited by the cost of the block-encoding step, which can dominate the overall resource requirements of a quantum algorithm, including circuit depth, gate counts, and qubit count \cite{Clader_2022}. Developing explicit and efficient block encodings is therefore central to making these algorithms practical.\\
In this work, we consider operators given as linear combinations of Pauli strings on \(n\) qubits with at most \(2n + 1\) terms, and we present a new block-encoding scheme for this setting, which we call Tag-And-Restore Encoding (TARE). TARE exploits a structural property of Pauli operators: A linear combination of pairwise anti-commuting Pauli strings with real coefficients whose squares sum to one is always unitary. This property has already been used in other quantum algorithms, for example in unitary partitioning for Variational Quantum Eigensolvers (VQE) to group Pauli terms into unitary subsets~\cite{PhysRevA.101.062322,doi:10.1021/acs.jctc.9b00791}, or in the construction of dressing Hamiltonians~\cite{lang2020unitarytransformationelectronichamiltonian, ryabinkin2023efficientconstructioninvolutorylinear}.\\
The core idea of TARE is to pair the target operator with a unitary that has the same coefficient magnitudes but whose Pauli strings are pairwise anti-commuting. The target Pauli strings themselves are not required to satisfy any commutation structure.  In contrast to LCU, the coefficients are not loaded into the ancilla register but are encoded in the system register through this accompanying unitary. We then compute transformations that connect the two operators and apply an additional operator inspired by quantum error correction codes \cite{Gottesman.1998,nielsen_chuang} to recover the original Pauli strings. This step introduces an ancilla register whose size scales logarithmically with the number of Pauli strings, constructs controlled tagging operations that correlate the anti-commuting Pauli strings with orthogonal ancilla states, and then applies the transformations controlled on those ancilla states to obtain a block encoding of the original operator. Through numerical simulations, we show that TARE substantially reduces the T-gate count compared to LCU, while the circuit depth can be lower than that of LCU for some examples.\\
TARE is not limited to logarithmic-size ancilla registers and can be easily extended to larger ancilla register sizes. This creates a trade-off between ancilla count and overall gate complexity. For example, we show that TARE can be extended to a linear-size ancilla register, which dramatically reduces the circuit depth.\\
This work is organized as follows. In Section~\ref{sec:notations}, we introduce the notations used throughout the paper. Section~\ref{sec:block_encoding} formally introduces the concept of block encoding and gives a brief overview of the LCU method. In Section~\ref{sec:main_method}, we define the operators considered in this work and outline the main idea of TARE, followed by a more detailed description including the formal proofs in Section~\ref{sec:details}. Finally, in Section~\ref{sec:num-comp}, we define three concrete examples, the transverse-field Ising model, the Jordan-Wigner image of a fermionic star Hamiltonian, and random Pauli-string operators, and use numerical simulations to compare the circuit depth and T-gate count to LCU.
\section{Notations and Conventions} 
\label{sec:notations}
In this work, we follow the standard conventions in the quantum computing literature where the Dirac notations $\bra{\cdot}$ and $\ket{\cdot}$ are used to denote row and column vectors, respectively. In particular, $\ket{0}$ and $\ket{1}$ are used to denote respectively the two canonical basis vectors $(1, 0)^T$ and $(0, 1)^T$ of $\mathbb{C}^2$.\\
For an $n$ qubit system, we number the individual qubits as $q_0,q_1,\dots,q_{n-1}$. In a quantum circuit diagram we number their wires from top to bottom and we abbreviate multi-wires with a slash. For example, for $n=3$ we have $q = q_0 q_1 q_2$ and
\begin{equation}
\begin{quantikz}
\lstick{$q$} &\qwbundle{}
\end{quantikz}
\quad=\quad
\begin{quantikz}
\lstick{$q_0$} & \\
\lstick{$q_1$} & \\
\lstick{$q_2$} &
\end{quantikz}
\ \ .
\end{equation}
We denote by $\mathbb{F}_2$ the finite field of two elements. Let $a \in \mathbb{N}$.
Every integer $v \in \{0, 1, \dots, 2^{a}-1\}$ has a unique binary representation $\bit{v} = \bit{v}_0 \bit{v}_1 \dots \bit{v}_{a-1} \in \mathbb{F}_2^a$ with the property
\begin{equation}
    v = \sum_{k=0}^{a-1} \bit{v}_k \, 2^{a-1-k} .
    \label{eq:binary-representation}
\end{equation}
We use the convention that an integer $v$ in $\ket{v}$ refers to its binary representation, i.e., $\ket{v} = \ket{\bit{v}_0 \bit{v}_1 \dots \bit{v}_{a-1}}$.
For two integers $v, w \in \{0, 1, \dots, 2^{a}-1\}$ we define
\begin{equation}
    \Delta(v, w)
    = \bit{v} \cdot \bit{w}
    = \sum_{k=0}^{a-1} \bit{v}_k \bit{w}_k ,
    \label{eq:delta}
\end{equation}
where $\bit{v}$ and $\bit{w}$ denote the binary representations of $v$ and $w$, respectively.\\
We denote the four single-qubit Pauli operators by
\begin{align}
I = \begin{pmatrix} 1 & 0 \\ 0 & 1 \end{pmatrix},\qquad
X = \begin{pmatrix} 0 & 1 \\ 1 & 0 \end{pmatrix},\qquad
Y = \begin{pmatrix} 0 & -\mathrm{i} \\ \mathrm{i} & 0 \end{pmatrix},\qquad
Z = \begin{pmatrix} 1 & 0 \\ 0 & -1 \end{pmatrix} .
\end{align}
Each Pauli operator can be identified with an element of $\mathbb{F}_2^2$ by
\begin{equation}
    I \leftrightarrow (0,0), \quad
    X \leftrightarrow (1,0), \quad
    Y \leftrightarrow (1,1), \quad
    Z \leftrightarrow (0,1) .
    \label{eq:pauli-to-binary}
\end{equation}
Additionally, the Hadamard gate plays a central role in this paper and is given by
\begin{equation}
   H = \frac{1}{\sqrt{2}}\begin{pmatrix} 1 & 1 \\ 1 & -1 \end{pmatrix}. 
\end{equation}
In the quantum circuit diagrams we will denote gates with rectangular boxes.
Furthermore, we denote the controlled version of a gate $U$ by $CU$ and visualize it in a quantum circuit diagram as
\begin{equation}
\begin{quantikz}
\lstick{$c$} & &\ctrl{1}&\\
\lstick{$q$} &\qwbundle{}& \gate{U}&
\end{quantikz}\quad.
\end{equation}
Here, $c$ is the control qubit and $q$ are the target qubits. The gate $U$ is only applied if qubit $c$ is in the state $\ket{1}$. In the case of several control qubits we use round boxes and the integer representation of the control state in the quantum circuit diagram.
For example, a controlled $U$ gate on two qubits with control state $\ket{2} = \ket{\bit{10}}$ is denoted by
\begin{equation}
\begin{quantikz}
\lstick{$c$} &\qwbundle{}&\gate[style={rounded corners}]{2} \vqw{1}&\\
\lstick{$q$} &\qwbundle{}&\gate{U}&
\end{quantikz}
\quad=\quad
\begin{quantikz}
\lstick{$c_0$}& &\ctrl{2}& \\
\lstick{$c_1$}& &\octrl{1}&\\
\lstick{$q$} &\qwbundle{}&\gate{U}&
\end{quantikz}\quad.
\end{equation}
%
%
We use the term Pauli string for a tensor product of single-qubit Pauli operators
\begin{equation}
    P
    = \bigotimes_{i=0}^{n-1} \sigma_i ,
    \qquad
    \sigma_i \in
    \bigl\{
    I, X, Y, Z
    \bigr\} ,
    \label{eq:definition-pauli-string}
\end{equation}
and by \emph{phased Pauli string} we mean an element of the form $\gamma P$ with $\gamma \in \{\pm 1, \pm \mathrm{i}\}$ and $P$ a Pauli string. Applying the per-qubit correspondence~\eqref{eq:pauli-to-binary} to each factor $\sigma_i$ of $P$, we obtain $(x_i, z_i) \in \mathbb{F}_2^2$ and map $P$ to a vector $p \in \mathbb{F}_2^{2n}$ as
\begin{equation}
    P \;\leftrightarrow\; (x_0, \dots, x_{n-1}, z_0, \dots, z_{n-1}) .
    \label{eq:pauli-to-symplectic}
\end{equation}
Note that the multiplication of two Pauli strings $P$ and $Q$ translates to the addition of their binary representation vectors,
\begin{equation}
    PQ \;\leftrightarrow\; p + q .
    \label{eq:pauli-multiplication-binary}
\end{equation}
In this paper, arithmetic on $\mathbb{F}_2^{2n}$ is always modulo $2$.
Let us define the symplectic product $\langle \cdot, \cdot \rangle_{\mathrm{symp}} \colon \mathbb{F}_2^{2n} \times \mathbb{F}_2^{2n} \to \mathbb{F}_2$ as
\begin{equation}
    \langle p, q \rangle_{\mathrm{symp}}
    = p^\top \Omega\, q ,
    \qquad
    \Omega
    = \begin{pmatrix} 0_n & I_n \\ I_n & 0_n \end{pmatrix} ,
    \label{eq:symplectic-form}
\end{equation}
where $I_n$ and $0_n$ are the $n \times n$ identity and zero matrices, respectively. With this, the commutation relation of two Pauli strings $P$ and $Q$ can be calculated as
\begin{subequations}
\begin{align}
    &\langle p, q \rangle_{\mathrm{symp}} = 0
    \;\Longleftrightarrow\;
    P \text{ and } Q \text{ commute},\\
    &\langle p, q \rangle_{\mathrm{symp}} = 1
    \;\Longleftrightarrow\;
    P \text{ and } Q \text{ anti-commute}.
    \label{eq:symplectic-commutation-rule}
\end{align}
Equivalently,
\begin{equation}
    Q P Q = (-1)^{\langle p, q \rangle_{\mathrm{symp}}}\, P .
    \label{eq:symplectic-conjugation-rule}
\end{equation}
\end{subequations}
Throughout this paper, we consider operators of the form
\begin{equation}
    A = \sum_{k=0}^{m-1} \alpha_k \, P_k ,
    \label{eq:weighted-sum-commuting-paulis}
\end{equation}
where $P_0, \dots, P_{m-1}$ are $n$-qubit Pauli strings and $\alpha_0, \dots, \alpha_{m-1} \in \mathbb{C}$ are complex coefficients. We write each coefficient in polar form with modulus $\rho_k$ and phase $\varphi_k$ as
\begin{equation}
    \alpha_k = \rho_k\,e^{\mathrm{i}\varphi_k}, \qquad \rho_k \geq 0, \quad \varphi_k \in [0, 2\pi) .
    \label{eq:polar-form}
\end{equation}
The Pauli strings $P_k$ and coefficients $\alpha_k$ are the primary input to the block encoding methods discussed in this paper. Since the operator $A$ in~\eqref{eq:weighted-sum-commuting-paulis} is only unitary in special cases \cite{PhysRevA.101.062322}, we generally cannot apply it directly to a quantum state and must instead embed it into a larger unitary matrix via a block encoding.
\section{Block Encoding}
\label{sec:block_encoding}
The technique of embedding a (non-unitary) matrix $A \in \mathbb{C}^{2^n \times 2^n}$ into a larger unitary matrix $U_A$, such that
\begin{subequations}
\label{eq:definition-block-encoding}
\begin{equation}
    U_A
    = \begin{pmatrix}
    A/\scalingFactorBE & * \\
    * & *
    \end{pmatrix}
    \in \mathbb{C}^{2^{n+a} \times 2^{n+a}}
    \label{eq:definition-block-encoding-1}
\end{equation}
is called block encoding. Here, $*$ denotes matrix blocks of appropriate size whose values are irrelevant, and $\scalingFactorBE \geq \|A\|_2$ is a sub-normalization scaling factor that ensures $\|A/\scalingFactorBE\|_2 \le 1$, which is required so that $U_A$ exists. In the following, we mark the quantum states of the $n$-qubit system register with $\ket{\cdot}_{\text{sys}}$ to distinguish them from the states of the $a$-qubit ancilla register $\ket{\cdot}_{\text{anc}}$.\\
For a composite state 
$\ket{w} = \ket{0}_{\text{anc}} \ket{\psi}_\text{sys}$
we have
\begin{equation}
    U_A \ket{w}
    = \ket{0}_{\text{anc}} \frac{1}{\scalingFactorBE} A
    \ket{\psi}_\text{sys}
    + \ket{\perp} \ ,
    \label{eq:definition-block-encoding-2}
\end{equation}
where $\ket{\perp}$ is orthogonal to
$\frac{1}{\scalingFactorBE}\ket{0}_{\text{anc}} A \ket{\psi}_\text{sys}$.
This means that if we measure the ancilla qubits as $\bit{0} \dots \bit{0}$, the system register contains $A \ket{\psi}_\text{sys} / \| A \ket{\psi}_\text{sys}\|_2$:
\begin{equation}
    \begin{quantikz}
    \lstick{$\ket{0}_{\text{anc}}$} &   \qwbundle{a}  & \gate[2]{U_A} &   \meter{\bit{0} \dots \bit{0}}\\
    \lstick{$\ket{\psi}_\text{sys}$} & \qwbundle{n}   &      & \rstick{$A \ket{\psi}_\text{sys} / \| A \ket{\psi}_\text{sys}\|_2$}
    \end{quantikz}\quad.
\end{equation}
\end{subequations}
The standard block encoding for operators of form~\eqref{eq:weighted-sum-commuting-paulis} is the Linear Combination of Unitaries (LCU) method \cite{Childs_2012}, which proceeds in three steps (see Figure~\ref{fig:lcu_circuit}).
\begin{figure}[h]
    \centering
    \begin{tikzpicture}
    \node[scale=0.88]{
    \begin{quantikz}
    \lstick{$\ket{0}_{\text{anc}}$} & \qwbundle{a} & \gate{\opPrepare} & \gate[style={rounded corners}]{0} \vqw{1}\gategroup[2,steps=4,style={dashed,rounded corners,inner sep=6pt},background,label style={label position=above}]{$\opSelect$} & \gate[style={rounded corners}]{1} \vqw{1} & \ \cdots\ & \gate[style={rounded corners}]{m{-}1} \vqw{1} & \gate{\opPrepare^\dagger} & \meter{\bit{0} \dots \bit{0}} \\
    \lstick{$\ket{\psi}_{\text{sys}}$} & \qwbundle{n} &  & \gate{e^{\mathrm{i}\varphi_0} P_0} & \gate{e^{\mathrm{i}\varphi_1} P_1} & \ \cdots\ & \gate{e^{\mathrm{i}\varphi_{m-1}} P_{m-1}} && \rstick{$\sim A\ket{\psi}_{\text{sys}}$}
    \end{quantikz}
    };
    \end{tikzpicture}
    \caption{LCU block encoding circuit: The ancillas are initialized in $\ket{0}_\text{anc}$, $\opPrepare$ encodes the coefficients $\alpha_k$, $\opSelect$ applies the Pauli strings $P_k$ controlled on the ancillas, and $\opPrepare^\dagger$ followed by post-selecting the ancilla measurement outcome $\bit{0} \dots \bit{0}$ yields a block encoding of $A$ with $\scalingFactorBE = \|\alpha\|_1$.}
    \label{fig:lcu_circuit}
\end{figure}
Starting from $\ket{0}_\text{anc}\ket{\psi}_\text{sys}$, the state evolution is given by
\begin{subequations}
\begin{align}
    \ket{0}_\text{anc}\ket{\psi}_\text{sys}
    &\xrightarrow{\opPrepare}& \hspace{-1cm}
    &\frac{1}{\sqrt{\|\alpha\|_1}} \sum_{k=0}^{m-1} \sqrt{\rho_k}\,\ket{k}_\text{anc}\ket{\psi}_\text{sys} \\
    &\xrightarrow{\opSelect}& \hspace{-1cm}
    &\frac{1}{\sqrt{\|\alpha\|_1}} \sum_{k=0}^{m-1} \sqrt{\rho_k}\,\ket{k}_\text{anc}\, e^{\mathrm{i}\varphi_k} P_k\ket{\psi}_\text{sys} \\
    &\xrightarrow{\opPrepare^\dagger}& \hspace{-1cm}
    &\frac{1}{\|\alpha\|_1} \ket{0}_\text{anc} \sum_{k=0}^{m-1} \rho_k\, e^{\mathrm{i}\varphi_k} P_k\ket{\psi}_\text{sys} + \ket{\perp} \notag \\ 
    && \hspace{-1cm} &= \frac{1}{\|\alpha\|_1} \ket{0}_\text{anc}\, A\ket{\psi}_\text{sys} + \ket{\perp} ,
\end{align}
\end{subequations}
where $\|\alpha\|_1 = \sum_{k=0}^{m-1} \rho_k$. This shows that the circuit is a block encoding of $A$ with subnormalization factor $\scalingFactorBE = \|\alpha\|_1$.
Note that the phases $e^{\mathrm{i}\varphi_k}$ are typically absorbed into the state preparation by $\opPrepare$. Here we assign them to $\opSelect$ to align with the structure of TARE in Section~\ref{sec:main_method}.
\section{Tag-And-Restore Encoding}
\label{sec:main_method}
In this section, we present our novel block encoding method for operator $A$ defined in~\eqref{eq:weighted-sum-commuting-paulis}. We call this method Tag-And-Restore Encoding (TARE), which will become clear in the course of this section. The method imposes no constraints on the Pauli strings $P_k$ or the coefficients $\alpha_k \in \mathbb{C}$. The only requirement is $m \leq 2n+1$, since our construction pairs the $P_k$ with a set of pairwise anti-commuting Pauli strings, which on $n$ qubits can have at most $2n+1$ elements \cite{sarkar2019setscommutinganticommutingpaulis}. Larger operators can still be handled by splitting them into groups of at most $2n+1$ strings and combining the resulting block encodings via an LCU step, but this is not further explored in this work.\\
We first give a high-level overview of the objects and circuit structure of TARE here and defer the formal derivation to Section~\ref{sec:details}.\\
Recalling the polar form~\eqref{eq:polar-form} of the coefficients, we first construct a unitary operator $U_\mathrm{anti}$ that uses the normalized moduli of the coefficients of $A$ in~\eqref{eq:weighted-sum-commuting-paulis} and the same number $m$ of Pauli strings, but replaces the $P_k$ with a family of pairwise anti-commuting Pauli strings $\Rop_k$,
\begin{subequations}
\begin{equation}
    U_\mathrm{anti}
    = \sum_{k=0}^{m-1} \frac{\rho_k}{\|\alpha\|_2} \, \Rop_k ,
    \qquad
    \Rop_i \Rop_j = - \Rop_j \Rop_i ,
    \ \ i \neq j,
    \label{eq:definition-U-main-text}
\end{equation}
where
\begin{equation}
    \|\alpha\|_2 = \sqrt{\sum_{k=0}^{m-1} \rho_k^2} .
    \label{eq:alpha-norm}
\end{equation}
\end{subequations}
Together with the anti-commutation of the $\Rop_k$, the normalization $\sum_k (\rho_k/\|\alpha\|_2)^2 = 1$ ensures that $U_\mathrm{anti}$ is always unitary \cite{PhysRevA.101.062322,doi:10.1021/acs.jctc.9b00791}. The phases $\varphi_k$ are absorbed into a later step.
The precise relation between the $P_k$ and the $\Rop_k$, and the freedom in choosing the $\Rop_k$, are made explicit in Section~\ref{sec:details}.\\
At the circuit level, $U_\mathrm{anti}$ is embedded into a larger system with $a$ ancilla qubits and pre- and appended with operations that turn the application of $U_\mathrm{anti}$ on the system qubits into an application of $A$. The corresponding quantum circuit is shown in Figure \ref{fig:full_circuit_compact}.
\begin{figure}[h]
     \centering
    \begin{tikzpicture}
    \node[scale=1] {
    \begin{quantikz}
    \lstick{$\ket{0}_{\text{anc}}$} &   \qwbundle{a}  &\gate[2,style={fill=red!20}]{\opTag}&      & \gate[2,style={fill=red!20}]{\opTag^\dagger} & \gate[2,style={fill=green!20}]{\opRestore} & \gate[style={fill=orange!0}]{W^\dagger}        & \meter{\bit{0} \dots \bit{0}}\\
    \lstick{$\ket{\psi}_\text{sys}$} & \qwbundle{n}   &      &\gate[1,style={fill=blue!20}]{U_\mathrm{anti}}&      &           & &\rstick{$ \sim A \ket{\psi}_\text{sys}$}
    \end{quantikz}
    };
    \end{tikzpicture}
    \caption{General quantum circuit structure of TARE.}
    \label{fig:full_circuit_compact}
\end{figure}
This circuit consists of three main steps. The first is the conjugation of $U_\mathrm{anti}$ with a unitary operator $\opTag$ that acts as
\begin{equation}
    \opTag^\dagger U_\mathrm{anti} \opTag \ket{0}_\text{anc}\ket{\psi}_\text{sys}
    =
    \sum_{k=0}^{m-1} \frac{\rho_k}{\|\alpha\|_2} \ket{c_k}_{\text{anc}} \Rop_k \ket{\psi}_\text{sys},
    \label{eq:U_tag_conjugation_U_A}
\end{equation}
where $c_k \in \{0, 1, \dots, 2^a - 1\}$ are integers such that $\ket{c_k}_{\text{anc}}$ are orthogonal control states of the ancilla register. This gives us control over the summands of $U_\mathrm{anti}$ so that in the second step the operator $\opRestore$ can transform each $\Rop_k$ into the corresponding $P_k$ while simultaneously applying the phase $e^{\mathrm{i}\varphi_k}$ on the branch labeled by $\ket{c_k}_\text{anc}$, i.e.,
\begin{equation}
    \opRestore\sum_{k=0}^{m-1}\frac{\rho_k}{\|\alpha\|_2}\ket{c_k}_{\text{anc}}\Rop_k\ket{\psi}_\text{sys}
    = \sum_{k=0}^{m-1} \frac{\alpha_k}{\|\alpha\|_2} \ket{c_k}_{\text{anc}} P_k\ket{\psi}_\text{sys} .
    \label{eq:U_restore}
\end{equation}
Note the similarity to the $\opSelect$ oracle of the LCU method.\\
For any positive integer $m$, any integer $a \geq \lceil \log_2 m \rceil$, and any distinct integers $c_0, \dots, c_{m-1} \in \{0, 1, \dots, 2^a - 1\}$, we denote by $W$ a unitary with
\begin{equation}
    W\ket{0}_\text{anc}
    = \frac{1}{\sqrt{m}} \sum_{k=0}^{m-1} \ket{c_k}_\text{anc} .
    \label{eq:W-definition}
\end{equation}
Applying $W^\dagger$ to the ancilla register as a third step gives the desired block encoding of $A$,
\begin{equation}
     (W^\dagger \otimes I^{\otimes n})
     \sum_{k=0}^{m-1} \frac{\alpha_k}{\|\alpha\|_2} \ket{c_k}_{\text{anc}} P_k  \ket{\psi}_\text{sys}
    = \frac{1}{\sqrt{m}\,\|\alpha\|_2} \ket{0}_{\text{anc}} A \ket{\psi}_\text{sys} + \ket{\perp},
\end{equation}
Comparing with~\eqref{eq:definition-block-encoding}, we see that the quantum circuit in Figure~\ref{fig:full_circuit_compact} is a block encoding of $A$ with
\begin{equation}
    \scalingFactorBE = \sqrt{m}\,\|\alpha\|_2 .
    \label{eq:lambda-BE-TARE}
\end{equation}
The LCU method applied to the same operator has $\scalingFactorBE_{\mathrm{LCU}} = \|\alpha\|_1$. The norm equivalence $\|\alpha\|_1 \leq \sqrt{m}\,\|\alpha\|_2$ gives
\begin{equation}
    \scalingFactorBE_{\mathrm{LCU}} \leq \scalingFactorBE ,
\end{equation}
with equality if and only if all coefficient moduli $|\alpha_k|$ are equal. TARE thus matches the subnormalization of LCU in the uniform-coefficient case and is strictly larger otherwise. The key advantage of TARE lies elsewhere: In Section~\ref{sec:num-comp} we will show that it can achieve a significant reduction in gate complexity.
\subsection{Details of Proposed Method}
\label{sec:details}
In this section, we present the formal results underlying TARE. The detailed circuit is given in Fig.~\ref{fig:general_quantum_circuit}.
\begin{figure}[h]
     \centering
    \begin{tikzpicture}
    \node[scale=0.55] {
    \begin{quantikz}[wire types={q,q,n,q,q,n}]
    \lstick{$\ket{0}$} &&\gate{H}\gategroup[5,steps=5,style={dashed,fill=red!20}, background]{\raisebox{0.2em}{$\opTag$}} &\ctrl{4}  &          &            &       &  &  \gategroup[5,steps=5,style={dashed,fill=red!20}, background]{\raisebox{0.2em}{$\opTag^\dagger$}}  &            &        & \ctrl{4}&\gate{H}&\gate[4,style={rounded corners}]{c_{0}}\gategroup[5,steps=4,style={dashed,fill=green!20}, background]{\raisebox{0.2em}{$\opRestore$}}&\gate[4,style={rounded corners}]{c_{1}}&\dots&\gate[4,style={rounded corners}]{c_{m-1}}&\gate[4,style={fill=orange!0}]{W^\dagger}&\meter{0}
    \\
    \lstick{$\ket{0}$} &&  \gate{H} &    &  \ctrl{3}&  &  &                   &        & &  \ctrl{3} &     &\gate{H}&&&\dots&&&\meter{0}
    \\
    \lstick{\vdots}
    \\
    \lstick{$\ket{0}$} &&\gate{H} &   &  & &\ctrl{1}     &      &       \ctrl{1}       &     &&   & \gate{H}&&&\dots&&&\meter{0}
    \\
    \lstick{$\ket{\psi}_\text{sys}$}  &  \qwbundle{n}        & \slice[label style={pos=1, anchor=north}]{\text{step 1}}&\gate{S_{0}}&\gate{S_{1}}&\dots& \gate{S_{a-1}}&   \gate[style={fill=blue!20}]{U_\mathrm{anti}}\slice[label style={pos=1, anchor=north}]{\text{step 2}}&\gate{S_{a-1}}&\dots&\gate{S_{1}}& \gate{S_{0}}\slice[label style={pos=1, anchor=north}]{\text{step 3}}&\slice[label style={pos=1, anchor=north}]{\text{step 4}}& \gate{e^{\mathrm{i}\varphi_0} T_0} \vqw{-1}&\gate{e^{\mathrm{i}\varphi_1} T_1} \vqw{-1}&\dots&\gate{e^{\mathrm{i}\varphi_{m-1}} T_{m-1}} \vqw{-1}\slice[label style={pos=1, anchor=north}]{\text{step 5}}&\slice[label style={pos=1, anchor=north}]{\text{step 6}}&\rstick{$ \sim A \ket{\psi}_\text{sys}$}
    \\
    &&&&&&&&&&&&&&&&&&&
    \end{quantikz}
    };
    \end{tikzpicture}
     \caption{Detailed quantum circuit of TARE with an arbitrary ancilla register size. The step labels (steps~1--6) correspond to the proof of Theorem~\ref{thm:main-thm}.}
     \label{fig:general_quantum_circuit}
 \end{figure}
%
\begin{theorem}
    \label{thm:main-thm}
    Let $a \geq \lceil \log_2 m \rceil$ be an integer and let $c_0, \dots, c_{m-1} \in \{0, 1, \dots, 2^a - 1\}$ be pairwise distinct. Let $S_0, \dots, S_{a-1}$ be Pauli strings and let $T_0, \dots, T_{m-1}$ be phased Pauli strings such that
    \begin{subequations}
    \begin{align}
        T_k \Rop_k &= P_k \ ,
        \qquad k = 0, \dots, m-1 \ , \quad \text{and}
        \label{eq:connection-Ptilde-R}
        \\
        \langle s_i, r_k \rangle_{\mathrm{symp}}
        &= \Delta(2^{a-1-i},\, c_k) \ ,
        \qquad k = 0, \dots, m-1 \ , \quad i = 0, \dots, a-1 \ ,
        \label{eq:thm-commutation-Si-Rk}
    \end{align}
    \end{subequations}
    where $P_0, \dots, P_{m-1}$ and $\Rop_0, \dots, \Rop_{m-1}$ are the Pauli strings of $A$ and $U_\mathrm{anti}$, respectively (see \eqref{eq:weighted-sum-commuting-paulis} and \eqref{eq:definition-U-main-text}), $s_0, \dots, s_{a-1}$ and $r_0, \dots, r_{m-1}$ are the symplectic vectors of $S_i$ and $\Rop_k$, respectively, and $W$ is given in~\eqref{eq:W-definition}. Then the quantum circuit in Fig.~\ref{fig:general_quantum_circuit} is a block encoding of $A$ with $\scalingFactorBE = \sqrt{m}\,\|\alpha\|_2$.
\end{theorem}
The proof uses the following Hadamard expansion identity. For $c \in \{0, 1, \dots, 2^{a}-1\}$,
\begin{equation}
    H^{\otimes a} \ket{c}
    = \tfrac{1}{\sqrt{2^a}}
    \sum_{\ell = 0}^{2^a - 1}
    \prod_{\substack{i=0 \\ \bit{l}_i=1}}^{a-1}
    (-1)^{\Delta(2^{a-1-i}, c)}
    \ket{\ell} \ .
    \label{eq:hadamard-tensor-a-ket-c}
\end{equation}
A formal statement and proof are given as Lemma~\ref{lemma:hadamard-expansion} in Appendix~\ref{Appendix:supporting-lemmas}.
\begin{proof}
    Going through the steps of the quantum circuit in Figure~\ref{fig:general_quantum_circuit} gives
    \begin{align*}
        \ket{0}_{\text{anc}}\ket{\psi}_{\text{sys}}
        \to&
        \tfrac{1}{\sqrt{2^a}}
        \sum_{\ell=0}^{2^a-1}
        \ket{\ell}_{\text{anc}}\ket{\psi}_{\text{sys}}
        && \text{\small step 1}
        \\
        \to&
        \tfrac{1}{\sqrt{2^a}}
        \sum_{\ell=0}^{2^a-1}
        \ket{\ell}_{\text{anc}}
        U_\mathrm{anti}
        \prod_{\substack{i=0 \\ \bit{l}_i=1}}^{a-1} S_i
        \ket{\psi}_{\text{sys}}
        && \text{\small step 2}
        \\
        &=
        \tfrac{1}{\sqrt{2^a}}
        \sum_{\ell=0}^{2^a-1}
        \ket{\ell}_{\text{anc}}
        \sum_{k=0}^{m-1} \frac{\rho_k}{\|\alpha\|_2} \Rop_k
        \prod_{\substack{i=0 \\ \bit{l}_i=1}}^{a-1} S_i
        \ket{\psi}_{\text{sys}}
        \\
        &\stackrel{\eqref{eq:symplectic-conjugation-rule}}{=}
        \tfrac{1}{\sqrt{2^a}}
        \sum_{\ell=0}^{2^a-1}
        \ket{\ell}_{\text{anc}}
        \sum_{k=0}^{m-1}
        \Bigl(
        \prod_{\substack{i=0 \\ \bit{l}_i=1}}^{a-1}
        (-1)^{\langle s_i, r_k \rangle_{\mathrm{symp}}} S_i
        \Bigr)
        \frac{\rho_k}{\|\alpha\|_2} \Rop_k
        \ket{\psi}_{\text{sys}}
        \\
        \to&
        \tfrac{1}{\sqrt{2^a}}
        \sum_{\ell=0}^{2^a-1}
        \ket{\ell}_{\text{anc}}
        \sum_{k=0}^{m-1}
        \Bigl(
        \prod_{\substack{i=0 \\ \bit{l}_i=1}}^{a-1}
        (-1)^{\langle s_i, r_k \rangle_{\mathrm{symp}}}
        \Bigr)
        \frac{\rho_k}{\|\alpha\|_2} \Rop_k
        \ket{\psi}_{\text{sys}}
        && \text{\small step 3}
        \\
        &\stackrel{\eqref{eq:thm-commutation-Si-Rk}}{=}
        \sum_{k=0}^{m-1}
        \tfrac{1}{\sqrt{2^a}}
        \sum_{\ell=0}^{2^a-1}
        \prod_{\substack{i=0 \\ \bit{l}_i=1}}^{a-1}
        (-1)^{\Delta(2^{a-1-i}, c_k)}
        \ket{\ell}_{\text{anc}}\,
        \frac{\rho_k}{\|\alpha\|_2} \Rop_k
        \ket{\psi}_{\text{sys}}
        \\
        &\stackrel{\eqref{eq:hadamard-tensor-a-ket-c}}{=}
        \sum_{k=0}^{m-1}
        H^{\otimes a}
        \ket{c_k}_{\text{anc}}\,
        \frac{\rho_k}{\|\alpha\|_2} \Rop_k
        \ket{\psi}_{\text{sys}}
        \\
        \to&
        \sum_{k=0}^{m-1}
        \ket{c_k}_{\text{anc}}\,
        \frac{\rho_k}{\|\alpha\|_2} \Rop_k
        \ket{\psi}_{\text{sys}}
        && \text{\small step 4}
    \end{align*}
    Applying steps~5--6 of Figure~\ref{fig:general_quantum_circuit} further evolves the state as
    \begin{align*}
        \sum_{k=0}^{m-1}
        \ket{c_k}_{\text{anc}}\,
        \frac{\rho_k}{\|\alpha\|_2} \Rop_k
        \ket{\psi}_\text{sys}
        \to&
        \sum_{k=0}^{m-1}
        \ket{c_k}_{\text{anc}}\,
        e^{\mathrm{i}\varphi_k} T_k\, \frac{\rho_k}{\|\alpha\|_2} \Rop_k
        \ket{\psi}_\text{sys}
        && \text{\small step 5}
        \\
        &\stackrel{\eqref{eq:connection-Ptilde-R}}{=}
        \frac{1}{\|\alpha\|_2}\sum_{k=0}^{m-1}
        \ket{c_k}_{\text{anc}}\,
        \alpha_k P_k
        \ket{\psi}_\text{sys}
        \\
        \to&
        \frac{1}{\|\alpha\|_2}\sum_{k=0}^{m-1}
        W^\dagger
        \ket{c_k}_{\text{anc}}\,
        \alpha_k P_k
        \ket{\psi}_\text{sys}
        && \text{\small step 6}
        \\
        &\stackrel{\eqref{eq:W-definition}}{=}
        \frac{1}{\sqrt{m}\,\|\alpha\|_2}
        \ket{0}_{\text{anc}}
        \sum_{k=0}^{m-1}
        \alpha_k P_k \ket{\psi}_\text{sys}
        + \ket{\perp} \ .
    \end{align*}
 This concludes the proof.
\end{proof}
\begin{remark}
\leavevmode
\begin{enumerate}[label=(\alph*)]
    \item The Pauli strings $\Rop_0, \dots, \Rop_{m-1}$, the number of ancillas $a$, and the control states $c_0, \dots, c_{m-1}$ in Theorem~\ref{thm:main-thm} can be chosen freely by the user. This enables tailoring the block encoding circuit to different constraints such as native hardware gates, circuit width, or depth.
    \item The Pauli strings $S_0, \dots, S_{a-1}$ can be computed from $\Rop_0, \dots, \Rop_{m-1}$ as we will detail in Theorem~\ref{thm:stabilizer-construction} below. As in~(a), they can be optimized to the user's need.
    \item $\opTag$ is a fully Clifford operation, as it consists solely of controlled Pauli strings, and therefore contributes no non-Clifford resources (e.g.\ T~gates) to the overall circuit cost.
\end{enumerate}
\end{remark}
\subsubsection{Implementation of $U_\mathrm{anti}$}
\label{sec:implementation}
One example of an explicit quantum circuit for $U_\mathrm{anti}$ is given by a product of $2m-1$ exponentials of Pauli terms~\cite{doi:10.1021/acs.jctc.9b00791},
\begin{subequations}
\begin{equation}
    U_\mathrm{anti} = -\mathrm{i} \prod_{j=0}^{m-2} \exp\left({\mathrm{i} \frac{\theta_{j}}{2} \Rop_j}\right) \exp\left({\mathrm{i}\, \theta_{m-1}\, \Rop_{m-1}}\right) \prod_{j=m-2}^0 \exp\left({\mathrm{i} \frac{\theta_{j}}{2} \Rop_j}\right),
    \label{eq:U_circuit}
\end{equation}
where the rotation angles $\theta_{j}$ are given by
\begin{equation}
    \theta_{j} = \arcsin\left( \frac{\rho_j}{\sqrt{\sum_{k=0}^{j} \rho_k^2}} \right).
\end{equation}
\end{subequations}
%
\subsubsection{Construction of Tagging Operations}
\label{sec:stabilizer-construction}
We now show that the tagging operations $S_0, \dots, S_{a-1}$ required by Theorem~\ref{thm:main-thm} can be constructed from any pairwise anti-commuting family $\Rop_0, \dots, \Rop_{m-1}$ with $m \le 2n{+}1$, under a mild consistency condition for odd $m$.

\begin{theorem}
\label{thm:stabilizer-construction}
Let $a, c_0, \dots, c_{m-1}, \Rop_0, \dots, \Rop_{m-1}$ be as in Theorem~\ref{thm:main-thm}.
If $m$ is odd and $\sum_{k=0}^{m-1} r_k = (0 \dots 0)^\top$, assume that the binary representations of $c_0, \dots, c_{m-1}$ sum to $\bit{0} \dots \bit{0}$,
\begin{equation}
    \bigoplus_{k=0}^{m-1} c_k = \bit{0} \dots \bit{0} \ .
    \label{eq:thm2-cond}
\end{equation}
Then, Pauli strings $S_0, \dots, S_{a-1}$ satisfying \eqref{eq:thm-commutation-Si-Rk} of Theorem~\ref{thm:main-thm} can be obtained as the solution to the linear system
\begin{subequations}
\begin{equation}
    \Pi^\top\, \Omega\, \Sigma = B \ ,
    \label{eq:linsys}
\end{equation}
where
\begin{equation}
\begin{aligned}
    \Pi &=
    \begin{pmatrix} | & & | \\ r_0 & \cdots & r_{m-1} \\ | & & | \end{pmatrix} ,
    &
    \Sigma &=
    \begin{pmatrix} | & & | \\ s_0 & \cdots & s_{a-1} \\ | & & | \end{pmatrix} ,
    \\[0.5em]
    B &=
    \begin{pmatrix} | & & | \\ b_0 & \cdots & b_{a-1} \\ | & & | \end{pmatrix} ,
    &
    b_i &=
    \begin{pmatrix} \Delta(2^{a-1-i}, c_0) \\ \vdots \\ \Delta(2^{a-1-i}, c_{m-1}) \end{pmatrix} .
\end{aligned}
\label{eq:linsys-def}
\end{equation}
\end{subequations}
The system \eqref{eq:linsys} is always solvable.
\end{theorem}

\begin{proof}
The linear system is directly obtained from \eqref{eq:thm-commutation-Si-Rk}, so only the solvability remains to be shown. Since $\Omega$ is invertible, it does not affect the solvability of \eqref{eq:linsys}. We distinguish two cases:
\begin{enumerate}[label=(\alph*)]
    \item If $m$ is even, or $m$ is odd with $\sum_k r_k \neq 0$, Lemma~\ref{lemma:linear-independence} in the appendix shows that $r_0, \dots, r_{m-1}$ are linearly independent. The matrix $\Pi$ thus has full column rank and \eqref{eq:linsys} is always solvable.
    \item If $m$ is odd and $\sum_k r_k = 0$, Lemma~\ref{lemma:linear-independence} in the appendix shows that $\Pi$ has rank $m-1$ with $\mathbf{1} = (1, \dots, 1)^\top$ spanning the kernel of $\Pi$. The consistency condition
    \begin{equation}
        \mathbf{1}^\top B = \bit{0} \dots \bit{0}        \label{eq:consistency}
    \end{equation}
    is therefore necessary and sufficient for the solvability of \eqref{eq:linsys}. By the definition of $B$, condition~\eqref{eq:consistency} is equivalent to
    \begin{equation*}
        \sum_{k=0}^{m-1} \Delta(2^{a-1-i}, c_k) = 0
        \qquad \text{for all } i = 0, \dots, a-1 \ .
    \end{equation*}
    By \eqref{eq:delta}, the left-hand side is bit $i$ of $\bigoplus_{k=0}^{m-1} c_k$. Since this holds for every $i = 0, \dots, a-1$, we obtain $\bigoplus_k c_k = \bit{0} \dots \bit{0}$, which is exactly condition~\eqref{eq:thm2-cond}.
\end{enumerate}
\end{proof}
\begin{remark}
\label{remark:s_i-optimization}
In general, the solution of \eqref{eq:linsys} is not unique. This can be exploited to tailor the choice of $S_0, \dots, S_{a-1}$ to different requirements. For example, \eqref{eq:linsys} could be solved for every individual $s_i$ such that the resulting $S_i$ has minimal weight. A potentially better strategy is to solve for all $\Sigma = (s_0, \dots, s_{a-1})$ simultaneously and optimize for the row-wise maximum $\|\Sigma\|_\infty$ to target circuit depth, or the total weight $\|\Sigma\|_T$ to target overall gate count. An explicit closed-form solution of \eqref{eq:linsys} is given in Appendix~\ref{Appendix:explicit-stabilizers}.
\end{remark}
\section{Numerical Comparison}
\label{sec:num-comp}
We compare LCU \cite{Childs_2012} and TARE on three examples. In all cases, the tagging operations $S_i$ are obtained from the linear system in Theorem~\ref{thm:stabilizer-construction}, with each $S_i$ independently optimized for minimal weight. The theorem leaves the control labels $c_k$ free up to a mild consistency condition for odd $m$, and we adopt the two choices $c_k = k$ for $a = \lceil \log_2 m \rceil$ and $c_k = 2^k$ for $a = m$. We fix the anti-commuting family
\begin{equation}
\Rop_{2j} = \left( \prod_{i=0}^{j-1} Z_i \right) X_j, \qquad
\Rop_{2j+1} = \left( \prod_{i=0}^{j-1} Z_i \right) Y_j, \qquad j=0,\dots,n-1,
\label{eq:anticomm_family}
\end{equation}
implement $U_\mathrm{anti}$ as in \eqref{eq:U_circuit}, and compute the transformation strings $T_k$ from \eqref{eq:connection-Ptilde-R}.\\
Both methods are implemented in Qrisp~\cite{seidel2024qrispframeworkcompilablehighlevel,petrič2026blockencodingsprogrammingabstractionseclipse}. The multi-controlled structure ($\opSelect$ in LCU and $\opRestore$ in TARE) is implemented by Qrisp's \texttt{q\_switch} with the \texttt{"tree"} method, which routes cases through a balanced binary tree~\cite{Khattar_2025}. This introduces workspace qubits for the intermediate control state, but since both routines use \texttt{q\_switch} with the same number of multi-controlled gates, the overhead is identical and does not bias the comparison.\\
We benchmark three configurations: LCU and the logarithmic TARE variant, which share $a = \lceil \log_2 m \rceil$ and $c_k = k$ (ignoring the \texttt{q\_switch} workspace), and a linear-ancilla TARE variant with $a = m$ and $c_k = 2^k$ (construction in Appendix~\ref{sec:linear-ancilla-appendix}). The operator $W$ then follows from the control labels: A uniform state over $\mathcal{O}(\log m)$ rotations~\cite{shukla2024efficientquantumcircuits} in the logarithmic case, and a Dicke-1 state from a chain of $m-1$ controlled rotations~\cite{Bartschi_2019} in the linear case.\\
We investigate the circuit depth and the T-gate count at synthesis precision $\varepsilon = 10^{-10}$, obtained with Qualtran~\cite{harrigan2024expressinganalyzingquantumalgorithms}. Qualtran decomposes the compiled circuit into Clifford+T with exact rotations, synthesising each non-Clifford rotation at $\lceil 1.149 \log_2(1/\varepsilon) + 9.2 \rceil$ T~gates.\\
The T-gate count is dominated by coefficient loading. In Qrisp's LCU, \opPrepare{} is realized via a binary tree of multiplexed $R_z$ rotations on the $a$ ancilla qubits. Each rotation is applied controlled during tree traversal, costing two non-Clifford $R_z$ rotations rather than one, and the coefficient array is zero-padded to $2^a$. For this reason the \opPrepare{} cost doubles whenever $m$ crosses a power-of-two boundary. TARE, in contrast, applies $2m-1$ uncontrolled Pauli rotations, growing strictly linearly in $m$.\\
The circuit depth is dominated by $\opSelect$ in LCU and $\opRestore$ in TARE, both scaling with the Pauli weights of $P_k$ and $T_k$ respectively. TARE therefore saves depth whenever the transformation strings $T_k$ have lower weight than the targets $P_k$.\\
The examples below use coefficients $\alpha_k$ drawn independently and uniformly at random. Here, TARE pays a small, $m$-independent overhead in subnormalization since the law of large numbers gives
\begin{equation}
    \frac{\scalingFactorBE_{\mathrm{LCU}}}{\scalingFactorBE}
    = \frac{\|\alpha\|_1}{\sqrt{m}\,\|\alpha\|_2}
    \xrightarrow{m \to \infty}
    \frac{\mathbb{E}[\,|\alpha_k|\,]}{\sqrt{\mathbb{E}[\,\alpha_k^2\,]}}
    = \frac{\sqrt{3}}{2},
    \label{eq:subnorm-ratio-uniform}
\end{equation}
with $\mathcal{O}(1/\sqrt{m})$ fluctuations, so TARE carries a subnormalization factor of roughly $2/\sqrt{3} \approx 1.155$ times that of LCU.
Weighting depth or T-gate count by subnormalization would merely rescale the TARE curves by this constant, so the advantages reported below (which exceed $1.155$) remain genuine after the correction.
\paragraph{Example 1: Transverse-Field Ising Model.}
\label{sec:example-ising}
As a first example, we consider the transverse-field Ising model (TFIM) on $n$ qubits with open boundary conditions
\begin{equation}
    \begin{aligned}
    A_{\mathrm{Ising}}
    &= \sum_{k=0}^{n-1} \beta_{X,k}\, X_k
    \ +\ \sum_{k=0}^{n-2} \beta_{Z,k}\, Z_k Z_{k+1} .
    \end{aligned}
    \label{eq:op_ising}
\end{equation}
The TFIM is a standard model in quantum many-body physics and quantum simulation.
We use random real coefficients $\beta_{X,k}, \beta_{Z,k}$ drawn uniformly (disordered Ising model). Note that $m = 2n-1 \le 2n + 1$, so the operator satisfies the constraint $m \le 2n+1$ required by TARE.\\
We transform $A_{\mathrm{Ising}}$ into the form~\eqref{eq:weighted-sum-commuting-paulis} by relabelling $\alpha_{2j} P_{2j} = \beta_{X,j} X_j$ and $\alpha_{2j+1} P_{2j+1} = \beta_{Z,j} Z_j Z_{j+1}$. The resulting transformations $T_k$ are
\begin{equation}
    T_{2j} = \prod_{i=0}^{j-1} Z_i , \qquad
    T_{2j+1} = (-i)\left(\prod_{i=0}^{j-1} Z_i\right) X_j\, Z_{j+1} ,
    \label{eq:ising_transformations}
\end{equation}
with $T_0 = I$.
Figure~\ref{fig:comparison_ising} shows that in the logarithmic-ancilla variant TARE attains a substantial T-gate reduction over LCU across all tested $n$. However, the circuit depth is larger for TARE since the $P_k$ have constant Pauli weight at most two while the $T_k$ grow with the system size, reaching Pauli weight $\mathcal{O}(n)$. The linear-ancilla variant shows depth and T-gate count very similar to LCU.
\begin{figure}[h]
    \centering
  \subfloat[]{%
     \scalebox{0.85}{
\begin{tikzpicture}[scale=1]

\definecolor{darkgrey176}{RGB}{176,176,176}
\definecolor{lightgrey204}{RGB}{204,204,204}
\definecolor{steelblue55126184}{RGB}{55,126,184}
\definecolor{crimson2282628}{RGB}{228,26,28}
\definecolor{mediumseagreen7717574}{RGB}{77,175,74}
\definecolor{darkorange25513718}{RGB}{255,137,18}

\begin{axis}[
legend cell align={left},
legend style={
  fill opacity=0.8,
  nodes={scale=0.8},
  draw opacity=1,
  text opacity=1,
  at={(0.03,0.97)},
  anchor=north west,
  draw=lightgrey204
},
tick align=outside,
tick pos=left,
x grid style={darkgrey176},
xlabel={Number of system qubits $n$},
xmajorgrids,
xmin=0, xmax=102,
xtick style={color=black},
y grid style={darkgrey176},
ylabel={Circuit depth},
ylabel style={xshift=-8pt},
ymajorgrids,
ymin=-473, ymax=16885,
ytick style={color=black},
scaled y ticks=base 10:-3,
ytick distance=1000
]
\addplot [semithick, steelblue55126184, mark=*, mark size=1.5, mark options={solid}]
table {%
5 184
10 381
15 510
20 779
25 942
30 1053
35 1483
40 1594
45 1757
50 1902
55 2030
60 2159
65 2878
70 2989
75 3135
80 3246
85 3425
90 3554
95 3682
100 3843
};
\addlegendentry{LCU}
\addplot [semithick, crimson2282628, mark=triangle*, mark size=2, mark options={solid}]
table {%
5 290
10 703
15 980
20 1671
25 2016
30 2452
35 3684
40 4218
45 4749
50 5355
55 6028
60 6741
65 9073
70 9897
75 10749
80 11693
85 12610
90 13629
95 14696
100 15781
};
\addlegendentry{TARE (log ancilla)}
\addplot [semithick, darkorange25513718, mark=square*, mark size=1.5, mark options={solid}]
table {%
5 149
10 328
15 509
20 701
25 867
30 1062
35 1246
40 1431
45 1599
50 1777
55 1978
60 2140
65 2364
70 2498
75 2703
80 2897
85 3032
90 3280
95 3421
100 3643
};
\addlegendentry{TARE (linear ancilla)}
\end{axis}

\end{tikzpicture}}}
    \hfill
  \subfloat[]{%
      \scalebox{0.85}{
\begin{tikzpicture}[scale=1]

\definecolor{darkgrey176}{RGB}{176,176,176}
\definecolor{lightgrey204}{RGB}{204,204,204}
\definecolor{steelblue55126184}{RGB}{55,126,184}
\definecolor{crimson2282628}{RGB}{228,26,28}
\definecolor{mediumseagreen7717574}{RGB}{77,175,74}
\definecolor{darkorange25513718}{RGB}{255,137,18}

\begin{axis}[
legend cell align={left},
legend style={
  fill opacity=0.8,
  nodes={scale=0.8},
  draw opacity=1,
  text opacity=1,
  at={(0.03,0.97)},
  anchor=north west,
  draw=lightgrey204
},
tick align=outside,
tick pos=left,
x grid style={darkgrey176},
xlabel={Number of system qubits $n$},
xmajorgrids,
xmin=0, xmax=102,
xtick style={color=black},
y grid style={darkgrey176},
ylabel={T-gate count ($\varepsilon = 10^{-10}$)},
ylabel style={xshift=-8pt},
ymajorgrids,
ymin=-1211, ymax=43224,
ytick style={color=black}
]
\addplot [semithick, steelblue55126184, mark=*, mark size=1.5, mark options={solid}]
table {%
5 1737
10 4146
15 4429
20 8100
25 8699
30 9546
35 16027
40 17256
45 17663
50 18346
55 17763
60 19488
65 40397
70 34620
75 35205
80 35184
85 34549
90 35794
95 36941
100 38210
};
\addlegendentry{LCU}
\addplot [semithick, crimson2282628, mark=triangle*, mark size=2, mark options={solid}]
table {%
5 914
10 2148
15 3212
20 4510
25 5366
30 6534
35 7960
40 9128
45 9984
50 10944
55 12008
60 13072
65 14962
70 16130
75 17090
80 18258
85 19010
90 20074
95 21138
100 21994
};
\addlegendentry{TARE (log ancilla)}
\addplot [semithick, darkorange25513718, mark=square*, mark size=1.5, mark options={solid}]
table {%
5 1394
10 3314
15 5234
20 7154
25 9074
30 10994
35 12914
40 14834
45 16754
50 18674
55 20594
60 22514
65 24434
70 26354
75 28274
80 30194
85 32114
90 34034
95 35954
100 37874
};
\addlegendentry{TARE (linear ancilla)}
\end{axis}

\end{tikzpicture}}}
 \vspace{-0.5cm}
  \caption{Numerical comparison of circuit depth and T-gate count ($\varepsilon = 10^{-10}$) for block encoding of the transverse-field Ising model $A_{\mathrm{Ising}}$ with $m = 2n-1$ terms.
  Colors distinguish methods: LCU (blue) with ancilla register size $a = \lceil \log_2 m \rceil$, TARE with ancilla register size $a = \lceil \log_2 m \rceil$ and binary control states $c_k = k$ (red), and TARE with ancilla register size $a = m$ and control states $c_k = 2^k$ (orange).}
  \label{fig:comparison_ising}
\end{figure}
\paragraph{Example 2: Fermionic Star Hamiltonian.}
\label{sec:example-star}
As a second example, we consider a fermionic star Hamiltonian, in which one central hub site couples to $n-1$ spoke sites through particle-number-conserving hopping,
\begin{equation}
    H_\star
    = \sum_{k=1}^{n-1} \beta_k \big( c_0^\dagger c_k + c_k^\dagger c_0 \big) .
    \label{eq:star_fermionic}
\end{equation}
%
Under the Jordan-Wigner mapping~\cite{JordanWigner1928}, the hopping between the hub and the $k$-th spoke acquires a nonlocal $Z_1 \cdots Z_{k-1}$ tail. The qubit image of~\eqref{eq:star_fermionic} reads
\begin{equation}
    A_\star
    = \frac{1}{2}\sum_{k=1}^{n-1} \beta_k
      \Big( X_0 Z_1 \cdots Z_{k-1} X_k
          + Y_0 Z_1 \cdots Z_{k-1} Y_k \Big) ,
    \label{eq:op_star}
\end{equation}
with real coefficients $\beta_k$ drawn uniformly. Note that $m = 2(n-1) \le 2n+1$, so the operator satisfies the constraint $m \le 2n+1$ required by TARE.\\
We transform $A_\star$ into the form~\eqref{eq:weighted-sum-commuting-paulis} by relabelling, for $j = 0, \dots, n-2$,
$\alpha_{2j} P_{2j} = \tfrac{1}{2} \beta_{j+1}\, X_0 Z_1 \cdots Z_j X_{j+1}$ and
$\alpha_{2j+1} P_{2j+1} = \tfrac{1}{2} \beta_{j+1}\, Y_0 Z_1 \cdots Z_j Y_{j+1}$.
With the anti-commuting family~\eqref{eq:anticomm_family}, the resulting transformations $T_k$ are
\begin{equation}
    \begin{aligned}
    T_0 &= X_1, &\qquad T_1 &= Y_1, \\
    T_{2j} &= Y_0\, Y_j\, X_{j+1}, &\qquad T_{2j+1} &= X_0\, X_j\, Y_{j+1} \qquad (j \ge 1),
    \end{aligned}
    \label{eq:star_transformations}
\end{equation}
each of constant Pauli weight at most three, independent of the system size. This example therefore realises the opposite regime to the TFIM: The target Pauli strings $P_k$ have weights up to $\mathcal{O}(n)$ while the $T_k$ remain of constant weight. Consistent with this, Figure~\ref{fig:comparison_star} shows that the logarithmic-ancilla TARE variant surpasses LCU in depth for $n \gtrsim 30$, and the gap widens with system size, while keeping the advantage in T-gate count for all $n$. The linear-ancilla variant achieves a better depth than LCU with a T-gate count very similar to it.
\begin{figure}[h]
    \centering
  \subfloat[]{%
     \scalebox{0.85}{
\begin{tikzpicture}[scale=1]

\definecolor{darkgrey176}{RGB}{176,176,176}
\definecolor{lightgrey204}{RGB}{204,204,204}
\definecolor{steelblue55126184}{RGB}{55,126,184}
\definecolor{crimson2282628}{RGB}{228,26,28}
\definecolor{mediumseagreen7717574}{RGB}{77,175,74}
\definecolor{darkorange25513718}{RGB}{255,137,18}

\begin{axis}[
legend cell align={left},
legend style={
  fill opacity=0.8,
  nodes={scale=0.8},
  draw opacity=1,
  text opacity=1,
  at={(0.03,0.97)},
  anchor=north west,
  draw=lightgrey204
},
tick align=outside,
tick pos=left,
x grid style={darkgrey176},
xlabel={Number of system qubits $n$},
xmajorgrids,
xmin=0, xmax=102,
xtick style={color=black},
y grid style={darkgrey176},
ylabel={Circuit depth},
ylabel style={xshift=-8pt},
ymajorgrids,
ymin=-409, ymax=14588,
ytick style={color=black},
scaled y ticks=base 10:-3,
ytick distance=1000
]
\addplot [semithick, steelblue55126184, mark=*, mark size=1.5, mark options={solid}]
table {%
5 116
10 452
15 679
20 1130
25 1456
30 1884
35 2612
40 3105
45 3649
50 4293
55 4919
60 5630
65 6356
70 7740
75 8584
80 9477
85 10437
90 11465
95 12491
100 13634
};
\addlegendentry{LCU}
\addplot [semithick, crimson2282628, mark=triangle*, mark size=2, mark options={solid}]
table {%
5 170
10 626
15 806
20 1344
25 1493
30 1675
35 2610
40 2791
45 2930
50 3079
55 3257
60 3413
65 3519
70 5324
75 5480
80 5659
85 5791
90 5950
95 6124
100 6258
};
\addlegendentry{TARE (log ancilla)}
\addplot [semithick, darkorange25513718, mark=square*, mark size=1.5, mark options={solid}]
table {%
5 135
10 330
15 525
20 720
25 915
30 1110
35 1305
40 1500
45 1695
50 1890
55 2085
60 2280
65 2475
70 2670
75 2865
80 3060
85 3255
90 3450
95 3645
100 3840
};
\addlegendentry{TARE (linear ancilla)}
\end{axis}

\end{tikzpicture}}}
    \hfill
  \subfloat[]{%
      \scalebox{0.85}{
\begin{tikzpicture}[scale=1]

\definecolor{darkgrey176}{RGB}{176,176,176}
\definecolor{lightgrey204}{RGB}{204,204,204}
\definecolor{steelblue55126184}{RGB}{55,126,184}
\definecolor{crimson2282628}{RGB}{228,26,28}
\definecolor{mediumseagreen7717574}{RGB}{77,175,74}
\definecolor{darkorange25513718}{RGB}{255,137,18}

\begin{axis}[
legend cell align={left},
legend style={
  fill opacity=0.8,
  nodes={scale=0.8},
  draw opacity=1,
  text opacity=1,
  at={(0.03,0.97)},
  anchor=north west,
  draw=lightgrey204
},
tick align=outside,
tick pos=left,
x grid style={darkgrey176},
xlabel={Number of system qubits $n$},
xmajorgrids,
xmin=0, xmax=102,
xtick style={color=black},
y grid style={darkgrey176},
ylabel={T-gate count ($\varepsilon = 10^{-10}$)},
ylabel style={xshift=-8pt},
ymajorgrids,
ymin=-1134, ymax=40468,
ytick style={color=black}
]
\addplot [semithick, steelblue55126184, mark=*, mark size=1.5, mark options={solid}]
table {%
5 920
10 3755
15 4408
20 8669
25 8650
30 9443
35 16102
40 16769
45 17628
50 19009
55 18032
60 19001
65 21284
70 33171
75 34126
80 36427
85 34898
90 34729
95 37592
100 37821
};
\addlegendentry{LCU}
\addplot [semithick, crimson2282628, mark=triangle*, mark size=2, mark options={solid}]
table {%
5 680
10 1956
15 3020
20 4318
25 5174
30 6342
35 7768
40 8936
45 9792
50 10752
55 11816
60 12880
65 13460
70 15938
75 16898
80 18066
85 18818
90 19882
95 20946
100 21802
};
\addlegendentry{TARE (log ancilla)}
\addplot [semithick, darkorange25513718, mark=square*, mark size=1.5, mark options={solid}]
table {%
5 1202
10 3122
15 5042
20 6962
25 8882
30 10802
35 12722
40 14642
45 16562
50 18482
55 20402
60 22322
65 24242
70 26162
75 28082
80 30002
85 31922
90 33842
95 35762
100 37682
};
\addlegendentry{TARE (linear ancilla)}
\end{axis}

\end{tikzpicture}}}
 \vspace{-0.5cm}
  \caption{Numerical comparison of circuit depth and T-gate count ($\varepsilon = 10^{-10}$) for block encoding of the fermionic star Hamiltonian $A_\star$ with $m = 2(n-1)$ terms.
  Colors distinguish methods: LCU (blue) with ancilla register size $a = \lceil \log_2 m \rceil$, TARE with ancilla register size $a = \lceil \log_2 m \rceil$ and binary control states $c_k = k$ (red), and TARE with ancilla register size $a = m$ and control states $c_k = 2^k$ (orange).}
  \label{fig:comparison_star}
\end{figure}
\paragraph{Example 3: Random Dense Pauli-String Operators.}
\label{sec:example-random}
For this example we randomly draw $m=n$ dense Pauli strings $P^{\mathrm{rand}}_0, \ldots, P^{\mathrm{rand}}_{m-1}$, where each qubit position is assigned one of the three non-identity Pauli labels $\{X, Y, Z\}$ independently with equal probability, so that every string has full Pauli weight $n$. With these we form the operator
\begin{equation}
    A_{\mathrm{rand}} = \sum_{k=0}^{m-1} \alpha_k P^{\mathrm{rand}}_k ,
    \label{eq:op_A_rand}
\end{equation}
using random real coefficients $\alpha_k$ drawn uniformly.
For each system size $n$, we generate a single random instance and report its circuit depth and T-gate count.
With $P_k$ and $T_k$ of comparable Pauli weight, Figure~\ref{fig:comparison_random} shows that the LCU and logarithmic-TARE depths stay close, converging to within a few percent of each other at $n=100$, while TARE retains a clear T-gate advantage. The linear-ancilla variant achieves a significantly better depth than LCU with a T-gate count very similar to it.
\begin{figure}[h]
    \centering
  \subfloat[]{%
     \scalebox{0.85}{
\begin{tikzpicture}[scale=1]

\definecolor{darkgrey176}{RGB}{176,176,176}
\definecolor{lightgrey204}{RGB}{204,204,204}
\definecolor{steelblue55126184}{RGB}{55,126,184}
\definecolor{crimson2282628}{RGB}{228,26,28}
\definecolor{mediumseagreen7717574}{RGB}{77,175,74}
\definecolor{darkorange25513718}{RGB}{255,137,18}

\begin{axis}[
legend cell align={left},
legend style={
  fill opacity=0.8,
  nodes={scale=0.8},
  draw opacity=1,
  text opacity=1,
  at={(0.03,0.97)},
  anchor=north west,
  draw=lightgrey204
},
tick align=outside,
tick pos=left,
x grid style={darkgrey176},
xlabel={Number of system qubits $n$},
xmajorgrids,
xmin=0, xmax=102,
xtick style={color=black},
y grid style={darkgrey176},
ylabel={Circuit depth},
ylabel style={xshift=-8pt},
ymajorgrids,
ymin=-361, ymax=12892,
ytick style={color=black},
scaled y ticks=base 10:-3,
ytick distance=1000
]
\addplot [semithick, steelblue55126184, mark=*, mark size=1.5, mark options={solid}]
table {%
5 100
10 271
15 438
20 753
25 1057
30 1367
35 1911
40 2321
45 2824
50 3369
55 3938
60 4565
65 5570
70 6280
75 7066
80 7876
85 8797
90 9724
95 10693
100 11754
};
\addlegendentry{LCU}
\addplot [semithick, crimson2282628, mark=triangle*, mark size=2, mark options={solid}]
table {%
5 141
10 377
15 596
20 981
25 1269
30 1615
35 2339
40 2740
45 3212
50 3719
55 4283
60 4848
65 6271
70 6970
75 7708
80 8457
85 9289
90 10226
95 11106
100 12049
};
\addlegendentry{TARE (log ancilla)}
\addplot [semithick, darkorange25513718, mark=square*, mark size=1.5, mark options={solid}]
table {%
5 85
10 192
15 290
20 391
25 497
30 595
35 697
40 804
45 903
50 1005
55 1111
60 1210
65 1317
70 1413
75 1526
80 1621
85 1728
90 1828
95 1933
100 2040
};
\addlegendentry{TARE (linear ancilla)}
\end{axis}

\end{tikzpicture}}}
    \hfill
  \subfloat[]{%
      \scalebox{0.85}{
\begin{tikzpicture}[scale=1]

\definecolor{darkgrey176}{RGB}{176,176,176}
\definecolor{lightgrey204}{RGB}{204,204,204}
\definecolor{steelblue55126184}{RGB}{55,126,184}
\definecolor{crimson2282628}{RGB}{228,26,28}
\definecolor{mediumseagreen7717574}{RGB}{77,175,74}
\definecolor{darkorange25513718}{RGB}{255,137,18}

\begin{axis}[
legend cell align={left},
legend style={
  fill opacity=0.8,
  nodes={scale=0.8},
  draw opacity=1,
  text opacity=1,
  at={(0.03,0.97)},
  anchor=north west,
  draw=lightgrey204
},
tick align=outside,
tick pos=left,
x grid style={darkgrey176},
xlabel={Number of system qubits $n$},
xmajorgrids,
xmin=0, xmax=102,
xtick style={color=black},
y grid style={darkgrey176},
ylabel={T-gate count ($\varepsilon = 10^{-10}$)},
ylabel style={xshift=-8pt},
ymajorgrids,
ymin=-581, ymax=20744,
ytick style={color=black}
]
\addplot [semithick, steelblue55126184, mark=*, mark size=1.5, mark options={solid}]
table {%
5 721
10 1737
15 1964
20 4050
25 3635
30 4815
35 8168
40 7524
45 7301
50 8413
55 9118
60 9448
65 19071
70 19387
75 16942
80 16968
85 16949
90 16991
95 18082
100 17864
};
\addlegendentry{LCU}
\addplot [semithick, crimson2282628, mark=triangle*, mark size=2, mark options={solid}]
table {%
5 444
10 1018
15 1698
20 2148
25 2724
30 3316
35 4022
40 4406
45 5086
50 5470
55 6150
60 6534
65 7472
70 8064
75 8640
80 8920
85 9600
90 10088
95 10768
100 10944
};
\addlegendentry{TARE (log ancilla)}
\addplot [semithick, darkorange25513718, mark=square*, mark size=1.5, mark options={solid}]
table {%
5 626
10 1586
15 2546
20 3506
25 4466
30 5426
35 6386
40 7346
45 8306
50 9266
55 10226
60 11186
65 12146
70 13106
75 14066
80 15026
85 15986
90 16946
95 17906
100 18866
};
\addlegendentry{TARE (linear ancilla)}
\end{axis}

\end{tikzpicture}}}
 \vspace{-0.5cm}
  \caption{Numerical comparison of circuit depth and T-gate count ($\varepsilon = 10^{-10}$) for block encoding of random Pauli-string operators $A_{\mathrm{rand}}$ with $m = n$ Pauli strings.
  Each data point represents a single random instance.
  Colors distinguish methods: LCU (blue) with ancilla register size $a = \lceil \log_2 m \rceil$, TARE with ancilla register size $a = \lceil \log_2 m \rceil$ and binary control states $c_k = k$ (red), and TARE with ancilla register size $a = m$ and control states $c_k = 2^k$ (orange).}
  \label{fig:comparison_random}
\end{figure}
\section{Conclusions}
In this work, we introduced a new block-encoding method for implementing linear combinations of Pauli strings. For a target operator \(A\) given by a linear combination of up to \(2n+1\) arbitrary Pauli strings on \(n\) system qubits with arbitrary complex coefficients, TARE constructs a corresponding unitary \(U_\mathrm{anti}\) from a freely chosen family of pairwise anti-commuting Pauli strings with matching coefficient moduli. By tagging these anti-commuting Pauli strings with orthogonal ancilla states and applying suitable transformations \(T_k\), the target operator \(A\) is recovered on the system register. This yields a block encoding for any operator of this form, using an ancilla register that scales logarithmically in the number of Pauli strings and that can be extended straightforwardly to larger sizes.\\
We evaluated the method for the transverse-field Ising model, the Jordan--Wigner image of a fermionic star Hamiltonian, and random Pauli-string operators, and compared its circuit depth and T-gate count with those of the LCU approach. Our numerical results show that TARE provides a substantial reduction in T-gate count compared to LCU, with the advantage increasing with system size. This reduction is particularly relevant in fault-tolerant settings, where T gates typically dominate the implementation cost. The logarithmic-ancilla version of TARE can also result in a lower circuit depth than LCU for some examples.\\
We also showed that increasing the ancilla register size can significantly reduce the circuit depth of TARE. In particular, for a linear ancilla register with \(a=m\) qubits, \(\opRestore\) reduces to single-qubit-controlled gates, leading to substantially lower depth than both LCU and the logarithmic-ancilla variant. While larger ancilla registers can also be used in LCU~\cite{dellachiara2025efficientlcublockencodings}, this requires redesigning the \(\opPrepare\) oracle, whereas TARE accommodates such extensions without modifying the coefficient-loading step \(U_\mathrm{anti}\). A systematic comparison across intermediate ancilla sizes remains an interesting direction for future work.\\
A fundamental limitation of the approach is that an \(n\)-qubit system admits at most \(2n+1\) pairwise anti-commuting Pauli strings~\cite{sarkar2019setscommutinganticommutingpaulis}. Operators with more terms can nevertheless be addressed by partitioning them into suitable subsets, constructing separate TARE block encodings, and combining these encodings using additional LCU steps. This provides a route toward implementing larger Hamiltonians relevant for practical quantum simulation problems.\\
A further efficiency improvement can arise when some target terms already coincide with the corresponding term of $U_\mathrm{anti}$, i.e. \(\alpha_k P_k = \rho_k \Rop_k\). In this case the corresponding transformation \(T_k\) is the identity and \(\varphi_k = 0\), so no tagging or restoration is needed for that term, and the effective number of tagging states reduces to the number of indices with \(\alpha_k P_k \neq \rho_k \Rop_k\), potentially shrinking the ancilla register and lowering the depth of \(\opTag\) and \(\opRestore\). Choosing the anti-commuting family \(\{\Rop_k\}\) to maximize such matches is therefore a promising direction for future work.\\
Our \opPrepare{} baseline is the recursive isometry decomposition of Iten et al.~\cite{Iten2016isometries} as implemented in Qrisp (via Qiskit~\cite{javadiabhari2024quantumcomputingqiskit}). More efficient constructions such as QROM/QROAM~\cite{Babbush_2018_qrom,Berry_2019_qroam} can substantially reduce the coefficient-loading cost, but require extra ancillas and incur alias-sampling error. A head-to-head comparison with these methods is an interesting direction for future work.\\
Finally, because the implementation of \(U_\mathrm{anti}\) in this work reduces to a staircase-structured cascade of Pauli rotations, recent parallelization techniques that compress such circuits from \(\mathcal{O}(n)\) to \(\mathcal{O}(\log n)\) depth could further reduce the overall depth of TARE~\cite{watts2025quantumprecomputation}.

\section*{Acknowledgments}
The authors thank Leon Rullkötter for carefully proofreading the manuscript and Thomas Wellens for valuable discussions.
\\
This work was funded by the Ministry of Economic Affairs, Labour, and Tourism Baden-W\"urttemberg within the Competence Center Quantum Computing Ba\-den-W\"urt\-tem\-berg.
\printbibliography

@article{Khattar_2025,
   title={Rise of conditionally clean ancillae for efficient quantum circuit constructions},
   volume={9},
   ISSN={2521-327X},
   url={http://dx.doi.org/10.22331/q-2025-05-21-1752},
   DOI={10.22331/q-2025-05-21-1752},
   journal={Quantum},
   publisher={Verein zur Forderung des Open Access Publizierens in den Quantenwissenschaften},
   author={Khattar, Tanuj and Gidney, Craig},
   year={2025},
   month=may, pages={1752}
}

@misc{seidel2024qrispframeworkcompilablehighlevel,
      title={Qrisp: A Framework for Compilable High-Level Programming of Gate-Based Quantum Computers},
      author={Raphael Seidel and Sebastian Bock and René Zander and Matic Petrič and Niklas Steinmann and Nikolay Tcholtchev and Manfred Hauswirth},
      year={2024},
      eprint={2406.14792},
      archivePrefix={arXiv},
      primaryClass={quant-ph},
      url={https://arxiv.org/abs/2406.14792},
}

@misc{petrič2026blockencodingsprogrammingabstractionseclipse,
      title={Block-encodings as programming abstractions: The Eclipse Qrisp BlockEncoding Interface},
      author={Matic Petrič and René Zander},
      year={2026},
      eprint={2604.18276},
      archivePrefix={arXiv},
      primaryClass={quant-ph},
      url={https://arxiv.org/abs/2604.18276},
}

@misc{harrigan2024expressinganalyzingquantumalgorithms,
      title={Expressing and Analyzing Quantum Algorithms with Qualtran}, 
      author={Matthew P. Harrigan and Tanuj Khattar and Charles Yuan and Anurudh Peduri and Noureldin Yosri and Fionn D. Malone and Ryan Babbush and Nicholas C. Rubin},
      year={2024},
      eprint={2409.04643},
      archivePrefix={arXiv},
      primaryClass={quant-ph},
      url={https://arxiv.org/abs/2409.04643}, 
}

@article{ryabinkin2023efficientconstructioninvolutorylinear,
author = {Ryabinkin, Ilya G. and Jena, Andrew J. and Genin, Scott N.},
title = {Efficient Construction of Involutory Linear Combinations of Anticommuting Pauli Generators for Large-Scale Iterative Qubit Coupled Cluster Calculations},
journal = {Journal of Chemical Theory and Computation},
volume = {19},
number = {6},
pages = {1722-1733},
year = {2023},
doi = {10.1021/acs.jctc.2c01155},
note = {PMID: 36820812},
URL = {https://doi.org/10.1021/acs.jctc.2c01155},
eprint = {https://doi.org/10.1021/acs.jctc.2c01155}
}

@article{liu2024efficientquantumcircuitblock,
title = {An efficient quantum circuit for block encoding a pairing Hamiltonian},
journal = {Journal of Computational Science},
volume = {85},
pages = {102480},
year = {2025},
issn = {1877-7503},
doi = {https://doi.org/10.1016/j.jocs.2024.102480},
url = {https://www.sciencedirect.com/science/article/pii/S1877750324002734},
author = {Diyi Liu and Weijie Du and Lin Lin and James P. Vary and Chao Yang}
}

@book{nielsen_chuang,
  author    = {Michael A. Nielsen and Isaac L. Chuang},
  title     = {Quantum computation and quantum information},
  publisher = {Cambridge University Press},
  year      = {2000},
  isbn      = {9780521635035}
}

@book{10.5555/1973124,
author = {Rieffel, Eleanor and Polak, Wolfgang},
title = {Quantum computing: a gentle introduction},
year = {2011},
isbn = {9780262015066},
publisher = {The MIT Press},
edition = {1st}
}

@article{doi:10.1021/acs.jctc.9b00791,
author = {Izmaylov, Artur F. and Yen, Tzu-Ching and Lang, Robert A. and Verteletskyi, Vladyslav},
title = {Unitary partitioning approach to the measurement problem in the Variational Quantum Eigensolver method},
journal = {Journal of Chemical Theory and Computation},
volume = {16},
number = {1},
pages = {190-195},
year = {2020},
doi = {10.1021/acs.jctc.9b00791},
URL = {https://doi.org/10.1021/acs.jctc.9b00791}}

@article{PhysRevA.101.062322,
  title = {Measurement reduction in variational quantum algorithms},
  author = {Zhao, Andrew and Tranter, Andrew and Kirby, William M. and Ung, Shu Fay and Miyake, Akimasa and Love, Peter J.},
  journal = {Phys. Rev. A},
  volume = {101},
  issue = {6},
  pages = {062322},
  numpages = {19},
  year = {2020},
  month = {Jun},
  publisher = {American Physical Society},
  doi = {10.1103/PhysRevA.101.062322},
  url = {https://link.aps.org/doi/10.1103/PhysRevA.101.062322}
}

@misc{dellachiara2025efficientlcublockencodings,
      title={Efficient LCU block encodings through Dicke states preparation}, 
      author={Filippo Della Chiara and Martina Nibbi and Yizhi Shen and Roel Van Beeumen},
      year={2025},
      eprint={2507.20887},
      archivePrefix={arXiv},
      primaryClass={quant-ph},
      url={https://arxiv.org/abs/2507.20887}, 
}

@misc{sturm2025efficientexplicitblockencoding,
      title={Efficient and explicit block encoding of finite difference discretizations of the Laplacian}, 
      author={Andreas Sturm and Niclas Schillo},
      year={2025},
      eprint={2509.02429},
      archivePrefix={arXiv},
      primaryClass={quant-ph},
      url={https://arxiv.org/abs/2509.02429}, 
}

@article{Childs_2012,
    title={Hamiltonian simulation using linear combinations of unitary operations},
    author={Andrew M. Childs and Nathan Wiebe},
    year={2012},
    volume={12},
    ISSN={1533-7146},
    url={http://dx.doi.org/10.26421/QIC12.11-12},
    DOI={10.26421/qic12.11-12},
    journal={Quantum Information and Computation},
    publisher={Rinton Press},
    month=nov }

@article{lang2020unitarytransformationelectronichamiltonian,
author = {Lang, Robert A. and Ryabinkin, Ilya G. and Izmaylov, Artur F.},
title = {Unitary Transformation of the Electronic Hamiltonian with an Exact Quadratic Truncation of the Baker-Campbell-Hausdorff Expansion},
journal = {Journal of Chemical Theory and Computation},
volume = {17},
number = {1},
pages = {66-78},
year = {2021},
doi = {10.1021/acs.jctc.0c00170},
note = {PMID: 33295175},
URL = {https://doi.org/10.1021/acs.jctc.0c00170},
eprint = {https://doi.org/10.1021/acs.jctc.0c00170}
}

@article{S_nderhauf_2024,
   title={Block-encoding structured matrices for data input in quantum computing},
   volume={8},
   ISSN={2521-327X},
   url={http://dx.doi.org/10.22331/q-2024-01-11-1226},
   DOI={10.22331/q-2024-01-11-1226},
   journal={Quantum},
   publisher={Verein zur Forderung des Open Access Publizierens in den Quantenwissenschaften},
   author={Sünderhauf, Christoph and Campbell, Earl and Camps, Joan},
   year={2024},
   month=jan, pages={1226} }

@article{doi:10.1137/22M1484298,
author = {Camps, Daan and Lin, Lin and Van Beeumen, Roel and Yang, Chao},
title = {Explicit quantum circuits for block encodings of certain sparse matrices},
journal = {SIAM Journal on Matrix Analysis and Applications},
volume = {45},
number = {1},
pages = {801-827},
year = {2024},
doi = {10.1137/22M1484298},
URL = {  https://doi.org/10.1137/22M1484298}}

@article{Leng_2025,
   title={Expanding hardware-efficiently manipulable Hilbert space via Hamiltonian embedding},
   volume={9},
   ISSN={2521-327X},
   url={http://dx.doi.org/10.22331/q-2025-09-11-1857},
   DOI={10.22331/q-2025-09-11-1857},
   journal={Quantum},
   publisher={Verein zur Forderung des Open Access Publizierens in den Quantenwissenschaften},
   author={Leng, Jiaqi and Li, Joseph and Peng, Yuxiang and Wu, Xiaodi},
   year={2025},
   month=sep, pages={1857} }

@misc{rullkötter2025resourceefficientvariationalblockencoding,
      title={Resource-efficient Variational Compilation of Block-Encodings},
      author={Leon Rullkötter and Sebastian Weber and Vamshi Mohan Katukuri and Christian Tutschku and Bharadwaj Chowdary Mummaneni},
      year={2025},
      eprint={2507.17658},
      archivePrefix={arXiv},
      primaryClass={quant-ph},
      url={https://arxiv.org/abs/2507.17658}, 
}

@article{Kikuchi_2023,
   title={Realization of quantum signal processing on a noisy quantum computer},
   volume={9},
   ISSN={2056-6387},
   url={http://dx.doi.org/10.1038/s41534-023-00762-0},
   DOI={10.1038/s41534-023-00762-0},
   number={1},
   journal={npj Quantum Information},
   publisher={Springer Science and Business Media LLC},
   author={Kikuchi, Yuta and Mc Keever, Conor and Coopmans, Luuk and Lubasch, Michael and Benedetti, Marcello},
   year={2023},
   month=sep }

@article{Low_2017,
   title={Optimal Hamiltonian simulation by quantum signal processing},
   volume={118},
   ISSN={1079-7114},
   url={http://dx.doi.org/10.1103/PhysRevLett.118.010501},
   DOI={10.1103/physrevlett.118.010501},
   number={1},
   journal={Physical Review Letters},
   publisher={American Physical Society (APS)},
   author={Low, Guang Hao and Chuang, Isaac L.},
   year={2017},
   month=jan }

@article{Low_2019,
   title={Hamiltonian simulation by qubitization},
   volume={3},
   ISSN={2521-327X},
   url={http://dx.doi.org/10.22331/q-2019-07-12-163},
   DOI={10.22331/q-2019-07-12-163},
   journal={Quantum},
   publisher={Verein zur Forderung des Open Access Publizierens in den Quantenwissenschaften},
   author={Low, Guang Hao and Chuang, Isaac L.},
   year={2019},
   month=jul, pages={163} }

@inproceedings{Gily_n_2019, series={STOC ’19},
   title={Quantum singular value transformation and beyond: exponential improvements for quantum matrix arithmetics},
   url={http://dx.doi.org/10.1145/3313276.3316366},
   DOI={10.1145/3313276.3316366},
   booktitle={Proceedings of the 51st Annual ACM SIGACT Symposium on Theory of Computing},
   publisher={ACM},
   author={Gilyén, András and Su, Yuan and Low, Guang Hao and Wiebe, Nathan},
   year={2019},
   month=jun, pages={193–204},
   collection={STOC ’19} }

@article{Martyn_2021,
   title={Grand unification of quantum algorithms},
   volume={2},
   ISSN={2691-3399},
   url={http://dx.doi.org/10.1103/PRXQuantum.2.040203},
   DOI={10.1103/prxquantum.2.040203},
   number={4},
   journal={PRX Quantum},
   publisher={American Physical Society (APS)},
   author={Martyn, John M. and Rossi, Zane M. and Tan, Andrew K. and Chuang, Isaac L.},
   year={2021},
   month=dec }

@article{Clader_2022,
   title={Quantum resources required to block-encode a matrix of classical data},
   volume={3},
   ISSN={2689-1808},
   url={http://dx.doi.org/10.1109/TQE.2022.3231194},
   DOI={10.1109/tqe.2022.3231194},
   journal={IEEE Transactions on Quantum Engineering},
   publisher={Institute of Electrical and Electronics Engineers (IEEE)},
   author={Clader, B. David and Dalzell, Alexander M. and Stamatopoulos, Nikitas and Salton, Grant and Berta, Mario and Zeng, William J.},
   year={2022},
   pages={1–23} }

@misc{sarkar2019setscommutinganticommutingpaulis,
      title={On sets of commuting and anticommuting Paulis}, 
      author={Rahul Sarkar and Ewout van den Berg},
      year={2019},
      eprint={1909.08123},
      archivePrefix={arXiv},
      primaryClass={quant-ph},
      url={https://arxiv.org/abs/1909.08123}, 
}

@article{Gottesman.1998,
 author = {Gottesman, Daniel},
 year = {1998},
 title = {Theory of fault-tolerant quantum computation},
 url = {http://arxiv.org/pdf/quant-ph/9702029},
 pages = {127--137},
 volume = {57},
 number = {1},
 issn = {1050-2947},
 journal = {Physical Review A},
 doi = {10.1103/PhysRevA.57.127},
}

@article{PRXQuantum.5.020368,
  title = {Generalized Quantum Signal Processing},
  author = {Motlagh, Danial and Wiebe, Nathan},
  journal = {PRX Quantum},
  volume = {5},
  issue = {2},
  pages = {020368},
  numpages = {16},
  year = {2024},
  month = {Jun},
  publisher = {American Physical Society},
  doi = {10.1103/PRXQuantum.5.020368},
  url = {https://link.aps.org/doi/10.1103/PRXQuantum.5.020368}
}

@misc{sünderhauf2023generalizedquantumsingularvalue,
      title={Generalized Quantum Singular Value Transformation},
      author={Christoph Sünderhauf},
      year={2023},
      eprint={2312.00723},
      archivePrefix={arXiv},
      primaryClass={quant-ph},
      url={https://arxiv.org/abs/2312.00723},
}

@misc{watts2025quantumprecomputation,
      title={Quantum precomputation: parallelizing cascade circuits and the {Moore}-{Nilsson} conjecture is false},
      author={Adam Bene Watts and Charles R. Chen and J. William Helton and Joseph Slote},
      year={2025},
      eprint={2510.04411},
      archivePrefix={arXiv},
      primaryClass={quant-ph},
      url={https://arxiv.org/abs/2510.04411},
}

@article{shukla2024efficientquantumcircuits,
      title={An efficient quantum algorithm for preparation of uniform quantum superposition states},
      author={Shukla, Alok and Vedula, Prakash},
      journal={Quantum Information Processing},
      volume={23},
      number={2},
      pages={38},
      year={2024},
      publisher={Springer},
      doi={10.1007/s11128-024-04258-4},
      url={https://doi.org/10.1007/s11128-024-04258-4},
}

@article{Bartschi_2019,
      title={Deterministic Preparation of {Dicke} States},
      author={B{\"a}rtschi, Andreas and Eidenbenz, Stephan},
      booktitle={Fundamentals of Computation Theory},
      series={Lecture Notes in Computer Science},
      volume={11651},
      pages={126--139},
      year={2019},
      publisher={Springer},
      doi={10.1007/978-3-030-25027-0_9},
      url={https://doi.org/10.1007/978-3-030-25027-0_9},
}

@article{JordanWigner1928,
      title={{\"U}ber das Paulische {\"A}quivalenzverbot},
      author={Jordan, P. and Wigner, E.},
      journal={Zeitschrift f{\"u}r Physik},
      volume={47},
      number={9},
      pages={631--651},
      year={1928},
      doi={10.1007/BF01331938},
}

@article{Iten2016isometries,
      title={Quantum circuits for isometries},
      author={Iten, Raban and Colbeck, Roger and Kukuljan, Ivan and Home, Jonathan and Christandl, Matthias},
      journal={Physical Review A},
      volume={93},
      number={3},
      pages={032318},
      year={2016},
      publisher={American Physical Society},
      doi={10.1103/PhysRevA.93.032318},
}

@misc{javadiabhari2024quantumcomputingqiskit,
      title={Quantum computing with Qiskit},
      author={Ali Javadi-Abhari and Matthew Treinish and Kevin Krsulich and Christopher J. Wood and Jake Lishman and Julien Gacon and Simon Martiel and Paul D. Nation and Lev S. Bishop and Andrew W. Cross and Blake R. Johnson and Jay M. Gambetta},
      year={2024},
      eprint={2405.08810},
      archivePrefix={arXiv},
      primaryClass={quant-ph},
      url={https://arxiv.org/abs/2405.08810},
}

@article{Berry_2019_qroam,
      title={Qubitization of Arbitrary Basis Quantum Chemistry Leveraging Sparsity and Low Rank Factorization},
      author={Berry, Dominic W. and Gidney, Craig and Motta, Mario and McClean, Jarrod R. and Babbush, Ryan},
      journal={Quantum},
      volume={3},
      pages={208},
      year={2019},
      doi={10.22331/q-2019-12-02-208},
}

@article{Babbush_2018_qrom,
      title={Encoding Electronic Spectra in Quantum Circuits with Linear T Complexity},
      author={Babbush, Ryan and Gidney, Craig and Berry, Dominic W. and Wiebe, Nathan and McClean, Jarrod and Paler, Alexandru and Fowler, Austin and Neven, Hartmut},
      journal={Phys. Rev. X},
      volume={8},
      pages={041015},
      year={2018},
      doi={10.1103/PhysRevX.8.041015},
}

@misc{boyd2024lowoverheadparallelisationlcu,
      title={Low-overhead parallelisation of LCU via commuting operators},
      author={Gregory Boyd},
      year={2024},
      eprint={2312.00696},
      archivePrefix={arXiv},
      primaryClass={quant-ph},
      url={https://arxiv.org/abs/2312.00696},
}

@article{Chakraborty_2024,
      title={Implementing any linear combination of unitaries on intermediate-term quantum computers},
      author={Chakraborty, Shantanav},
      journal={Quantum},
      volume={8},
      pages={1496},
      year={2024},
      publisher={Verein zur F\"orderung des Open Access Publizierens in den Quantenwissenschaften},
      doi={10.22331/q-2024-10-10-1496},
      url={https://doi.org/10.22331/q-2024-10-10-1496},
}

@inproceedings{Camps_2022_FABLE,
      title={{FABLE}: Fast approximate quantum circuits for block-encodings},
      author={Camps, Daan and Van Beeumen, Roel},
      booktitle={2022 IEEE International Conference on Quantum Computing and Engineering (QCE)},
      pages={104--113},
      year={2022},
      publisher={IEEE},
      doi={10.1109/QCE53715.2022.00029},
}

@article{Loaiza_2025_Majorana,
      title={Majorana tensor decomposition: a unifying framework for decompositions of fermionic Hamiltonians to linear combination of unitaries},
      author={Loaiza, Ignacio and Sankar Brahmachari, Aritra and Izmaylov, Artur F.},
      journal={Quantum Science and Technology},
      volume={10},
      number={3},
      pages={035035},
      year={2025},
      publisher={IOP Publishing},
      doi={10.1088/2058-9565/add9c1},
}
\onecolumn\newpage
\appendix

\section{Linear Ancilla Register Variant}
\label{sec:linear-ancilla-appendix}
TARE extends naturally to a linear ancilla register of $a = m$ qubits. Choosing the control states $c_k$ (see Figure~\ref{fig:general_quantum_circuit}) as
\begin{equation}
    c_k = 2^k = (\underbrace{0 \dots 0}_{m-1-k} 1 \underbrace{0 \dots 0}_{k}),
        \qquad
        k = 0, \dots, m-1,
        \label{eq:control-states-lin-ancilla-register}
\end{equation}
makes each conditional branch of $\opRestore$ a single-qubit-controlled gate, since exactly one ancilla qubit is set for each $c_k$. The resulting circuit is shown in Figure~\ref{fig:restore_lin}. This removes the multi-controlled structure of $\opRestore$ and therefore drastically reduces its depth, at the cost of $m - \lceil \log_2 m \rceil$ additional ancilla qubits. The subnormalization factor $\scalingFactorBE = \sqrt{m}\,\|\alpha\|_2$ is unchanged from the logarithmic variant.
\begin{figure}[h]
     \centering
    \begin{tikzpicture}[baseline=(current bounding box.center)]
    \node[scale=0.75] {
    \begin{quantikz}
    \lstick{${\ell}$} & \qwbundle{a} & \gate[style={rounded corners}]{c_0}\gategroup[2,steps=4,style={dashed,fill=green!20}, background]{$\opRestore$} & \gate[style={rounded corners}]{c_1} & \dots & \gate[style={rounded corners}]{c_{m-1}} & \\
    \lstick{${q}$} & \qwbundle{n} & \gate{e^{\mathrm{i}\varphi_0} T_0}\vqw{-1} & \gate{e^{\mathrm{i}\varphi_1} T_1}\vqw{-1} & \dots & \gate{e^{\mathrm{i}\varphi_{m-1}} T_{m-1}}\vqw{-1} &
    \end{quantikz}
    };
    \end{tikzpicture}%
    $=$%
    \begin{tikzpicture}[baseline=(current bounding box.center)]
    \node[scale=0.75] {
    \begin{quantikz}[wire types={q,q,n,q,q}]
    \lstick{${\ell_0}$}  &        &\gategroup[5,steps=4,style={dashed,fill=green!20}, background]{$\opRestore$} & &&\ctrl{4}&  \\
    \lstick{${\ell_1}$}  &     & & & \ctrl{3} & &\\
    \lstick{\vdots}&&&&&\\
    \lstick{${\ell_{m-1}}$} &     & \ctrl{1} &&&  &\\
    \lstick{${q}$} &\qwbundle{n}&\gate{e^{\mathrm{i}\varphi_0} T_0}      &  \dots  &  \gate{e^{\mathrm{i}\varphi_{m-2}} T_{m-2}} & \gate{e^{\mathrm{i}\varphi_{m-1}} T_{m-1}}   &
    \end{quantikz}
    };
    \end{tikzpicture}
    \caption{Quantum circuit of $\opRestore$ in its general form (left) and its reduction under $a = m$ and $c_k = 2^k$ to single-qubit-controlled gates (right).}
    \label{fig:restore_lin}
\end{figure}

\section{Auxiliary Lemmas}
\label{Appendix:supporting-lemmas}
This appendix collects technical results referenced by the main text.\\
\begin{lemma}
\label{lemma:hadamard-expansion}
    For $c \in \{0, 1, \dots, 2^{a}-1\}$, the Hadamard expansion identity~\eqref{eq:hadamard-tensor-a-ket-c} holds, i.e.,
    \begin{equation*}
        H^{\otimes a} \ket{c}
        = \tfrac{1}{\sqrt{2^a}}
        \sum_{\ell = 0}^{2^a - 1}
        \prod_{\substack{i=0 \\ \bit{l}_i=1}}^{a-1}
        (-1)^{\Delta(2^{a-1-i}, c)}
        \ket{\ell} \ .
    \end{equation*}
\end{lemma}
\begin{proof}
    To prove~\eqref{eq:hadamard-tensor-a-ket-c}, we start from the definition of the Hadamard gate $H$, which gives
    \begin{equation}
        H^{\otimes a} \ket{c}
        = \tfrac{1}{\sqrt{2^a}}
        \sum_{\ell = 0}^{2^a - 1}
        (-1)^{\Delta(\ell, c)} \ket{\ell} \ .
        \label{eq:hadamard-tensor-a}
    \end{equation}
    The binary representation of an integer $\ell \in \{0, 1, \dots, 2^a-1\}$ can be written as
    \begin{equation}
        \bit{l}
        = \bigoplus_{\substack{i=0 \\ \bit{l}_i=1}}^{a-1}
        \bit{2}^{\bit{a-1-i}} \ ,
        \qquad
        \bit{2}^{\bit{a-1-i}}
        =
        \underbrace{\bit{0} \dots \bit{0}}_{i}
        \bit{1}
        \underbrace{\bit{0} \dots \bit{0}}_{a-1-i} \ ,
    \end{equation}
    where $\oplus$ is the bitwise addition modulo 2. In \cite{10.5555/1973124} it is shown that
    for integers $x, y,$ and $z$ and associated bitstrings $\bit{x}, \bit{y},$ and $\bit{z}$ it holds
    \begin{equation}
        (\bit{x} \oplus  \bit{y}) \cdot \bit{z}
        = \Delta(x, z) + \Delta(y, z) \ ,
        \label{eq:bitstring-integer-addition-equality}
    \end{equation}
    Combining the last two equations we obtain
    \begin{equation}
        \Delta(\ell, c)
        =
        \sum_{\substack{i=0 \\ \bit{l}_i=1}}^{a-1}
        \Delta(2^{a-1-i}, c) \ .
    \end{equation}
    Inserting this in \eqref{eq:hadamard-tensor-a} concludes the proof.
\end{proof}
%
\begin{lemma}
\label{lemma:linear-independence}
Let $R_0, \ldots, R_{m-1}$ be pairwise anti-commuting Pauli strings and let
$r_0, \ldots, r_{m-1} \in \mathbb{F}_2^{2n}$ be their symplectic vectors. Let
$\upsilon_0, \ldots, \upsilon_{m-1} \in \mathbb{F}_2$ such that
\begin{equation}
    \sum_{k=0}^{m-1} \upsilon_k r_k = 0 .
    \label{eq:lem-linindep-hyp}
\end{equation}
Then we have
\begin{equation}
    \upsilon_0 = \upsilon_1 = \cdots = \upsilon_{m-1} = \upsilon
    \label{eq:lem-linindep-conclusion}
\end{equation}
for a $\upsilon \in \mathbb{F}_2$. If $m$ is even, then $\upsilon = 0$.
\end{lemma}

\begin{proof}
Taking the symplectic inner product of \eqref{eq:lem-linindep-hyp} with $r_j$ gives
\begin{equation}
    0 = \sum_{k=0}^{m-1} \upsilon_k \langle r_k, r_j \rangle_{\mathrm{symp}}
      = \sum_{\substack{k=0 \\ k \neq j}}^{m-1} \upsilon_k .
    \label{eq:lem-linindep-innerprod}
\end{equation}
Here we used that the anti-commutation of $R_0, \ldots, R_{m-1}$ implies
$\langle r_k, r_j \rangle_{\mathrm{symp}} = 1$ for all $k \neq j$, and that
$\langle r_k, r_k \rangle_{\mathrm{symp}} = 0$.

Subtracting \eqref{eq:lem-linindep-innerprod} for two indices $j \neq j'$ we obtain
\[
    0 = \upsilon_j + \upsilon_{j'}  \qquad \text{for all } j,j' \in \{0,1,\ldots,m-1\},\; j \neq j'.
\]
Consequently, all $\upsilon_k$ must have the same value $\upsilon_k = \upsilon \in \mathbb{F}_2$,
proving \eqref{eq:lem-linindep-conclusion}. Inserting \eqref{eq:lem-linindep-conclusion} into \eqref{eq:lem-linindep-innerprod} yields
$(m-1)\, \upsilon = 0$, and thus $\upsilon = 0$ if $m$ is even.
\end{proof}
%
\section{Explicit Closed-Form Tagging Operations}
\label{Appendix:explicit-stabilizers}
Theorem~\ref{thm:stabilizer-construction} establishes that the linear system~\eqref{eq:linsys} is always solvable but does not provide an explicit solution. The following lemma gives a closed-form expression for the coefficients $\upsilon_{ki}$ in the ansatz $s_i = \sum_k \upsilon_{ki}\, r_k$, with separate formulas for even and odd $m$.
\begin{lemma}
\label{lemma:explicit-stabilizers}
Let $a, c_0, \dots, c_{m-1}, \Rop_0, \dots, \Rop_{m-1}$ be as in 
Theorem~\ref{thm:main-thm}. If $m$ is odd, assume that
$\bigoplus_{k=0}^{m-1} c_k = \bit{0} \dots \bit{0}$. Then,
the Pauli strings $S_0, \dots, S_{a-1}$ satisfying
\eqref{eq:thm-commutation-Si-Rk}
can be constructed from $\Rop_0, \dots, \Rop_{m-1}$ via
\begin{subequations}
\label{eq:explicit-stabilizers}
\begin{equation}
    s_i \;=\; \sum_{k=0}^{m-1} \upsilon_{ki}\, r_k \ ,
    \qquad \upsilon_{ki} \in \mathbb{F}_2 \ .
    \label{eq:stabilizer-ansatz}
\end{equation}
If $m$ is even, the coefficients $\upsilon_{ki}$ are given by
\begin{equation}
    \upsilon_{ki} \;=\;
    \sum_{\substack{j=0 \\ j\neq k}}^{m-1} \Delta(2^{a-1-i}, c_j) \ ,
    \label{eq:mu-even}
\end{equation}
and if $m$ is odd by
\begin{equation}
    \upsilon_{ki} \;=\; \Delta(2^{a-1-i}, c_k) \ .
    \label{eq:mu-odd}
\end{equation}
\end{subequations}
\end{lemma}
\begin{proof}
    Using the notation \eqref{eq:linsys-def} we can write
    \eqref{eq:stabilizer-ansatz} as
    \begin{equation}
        \Sigma = \Pi\, \Upsilon \ ,
        \qquad
        \Upsilon = (\upsilon_{ki}) \in \mathbb{F}_2^{m \times a} \ .
        \label{eq:proof-lemma-explicit-stabilizers-1}
    \end{equation}
    Inserting this into
    \eqref{eq:linsys}
    we obtain
    \begin{equation}
        \Theta \Upsilon = B \ ,
        \qquad \text{where }
        \Theta
        = \Pi^T \Omega\, \Pi
        {%
        \setlength{\arraycolsep}{2pt}%
        \renewcommand{\arraystretch}{0.5}%
        = \begin{pmatrix}
            0 & 1 & \cdots & \cdots & 1 \\[0.5pt]
            1 & \ddots & \ddots & & \vdots \\[0.5pt]
            \vdots & \ddots & \ddots & \ddots & \vdots \\[0.5pt]
            \vdots & & \ddots & \ddots & 1 \\[0.5pt]
            1 & \cdots & \cdots & 1 & 0
        \end{pmatrix}%
        } \ ,
        \label{eq:proof-lemma-explicit-stabilizers-2}
    \end{equation}
    and $B$ is given in \eqref{eq:linsys-def}. A direct calculation gives
    $(\Theta^2)_{j,l} = \sum_{k=0}^{m-1}(1-\delta_{jk})(1-\delta_{kl}) = m - 2 + \delta_{jl}$.
    If $m$ is even, $m - 2 + \delta_{jl} = \delta_{jl}$, so $\Theta^2 = I$ and the solution of
    \eqref{eq:proof-lemma-explicit-stabilizers-2}
    is
    \begin{equation}
        \Upsilon
        = \Theta B
        = \Bigl(
        \sum_{\substack{j=0 \\ j\neq k}}^{m-1} \Delta(2^{a-1-i}, c_j)
        \Bigr)_{k,i} \ .
    \end{equation}
    Inserting this into \eqref{eq:proof-lemma-explicit-stabilizers-1}
    yields \eqref{eq:mu-even}. 
    If $m$ is odd, we calculate $\Theta \Upsilon$ for $\Upsilon$ given by 
    \eqref{eq:mu-odd}:
    \begin{equation}
    \begin{aligned}
        \bigl(\Theta \Upsilon\bigr)_{k,i}
        = \sum_{\substack{j=0 \\ j\neq k}}^{m-1} \upsilon_{ji}
        &= \sum_{\substack{j=0 \\ j\neq k}}^{m-1} \Delta(2^{a-1-i}, c_j)
        \\
        &= \sum_{j=0}^{m-1} \Delta(2^{a-1-i}, c_j)
        + \Delta(2^{a-1-i}, c_k)
        = \Delta(2^{a-1-i}, c_k)
        = B_{k,i} \ .
    \end{aligned}
    \end{equation}
    Here, we used that the 
    condition $\bigoplus_{k=0}^{m-1} c_k = \bit{0} \dots \bit{0}$ results in
    \begin{equation}
        \sum_{j=0}^{m-1} \Delta(2^{a-1-i}, c_j)  = 0 \ ,
    \end{equation}
    see the proof of Theorem~\ref{thm:stabilizer-construction}.
    Thus, \eqref{eq:mu-odd} is a solution of 
    \eqref{eq:proof-lemma-explicit-stabilizers-2}, which concludes the proof.
\end{proof}
\end{document}